\documentclass[preprint,3p,times]{elsarticle}
\usepackage{amsmath,amssymb}
\usepackage[T1]{fontenc}
\usepackage[utf8]{inputenc}
\usepackage{graphicx}
\usepackage{booktabs}
\usepackage{siunitx}
\DeclareSIUnit\molar{M}
\DeclareSIUnit\ppm{ppm}
\DeclareSIUnit\dBm{dBm}
\usepackage{chemformula}
\usepackage{textcomp}
\usepackage{url}
\usepackage{xcolor}

\graphicspath{{./}}

\journal{Carbon}

\begin{document}

\begin{frontmatter}

\title{Solution-phase fluorination of nanodiamond: near-surface NV$^-$ activation and spin relaxation}

\author[obudaDS,wigner]{Szabolcs Czene}
\author[bme,cer]{Olga Krafcsik}
\author[wigner]{Nikoletta Jegenyés}
\author[wigner,obudaK]{Dávid Beke}
\author[wigner]{László Péter}
\author[wigner]{Vladimir Verkhovlyuk}
\author[wigner,bme,lend]{Ádám Gali\corref{cor}}
\ead{gali.adam@wigner.hun-ren.hu}
\cortext[cor]{Corresponding author.}

\affiliation[obudaDS]{organization={Doctoral School on Materials Sciences and Technologies, Óbuda University},
  addressline={Bécsi út 96/b}, postcode={H-1034}, city={Budapest}, country={Hungary}}
\affiliation[wigner]{organization={HUN-REN Wigner Research Centre for Physics, Institute for Solid State Physics and Optics},
  addressline={P.O. Box 49}, postcode={H-1525}, city={Budapest}, country={Hungary}}
\affiliation[bme]{organization={Department of Atomic Physics, Institute of Physics, Budapest University of Technology and Economics},
  addressline={Műegyetem rkp. 3.}, postcode={H-1111}, city={Budapest}, country={Hungary}}
\affiliation[cer]{organization={HUN-REN Centre for Energy Research, Institute of Technical Physics and Materials Science},
  addressline={P.O. Box 49}, postcode={H-1525}, city={Budapest}, country={Hungary}}
\affiliation[obudaK]{organization={Kandó Kálmán Faculty of Electrical Engineering, Óbuda University},
  addressline={Bécsi út 94-96}, postcode={H-1034}, city={Budapest}, country={Hungary}}
\affiliation[lend]{organization={MTA-WFK Lendület ``Momentum'' Semiconductor Nanoparticles Research Group},
  addressline={P.O. Box 49}, postcode={H-1525}, city={Budapest}, country={Hungary}}

\begin{abstract}
Nanodiamonds hosting luminescent point defects, known as fluorescent nanodiamonds (FND), are a leading platform for quantum technology. The nitrogen--vacancy (NV) centre is the most intensively studied of these; its negatively charged state (\ch{NV-}) can serve as a qubit at room temperature, and its stability is governed by surface functional groups. We present two solution-phase fluorination routes for stabilising \ch{NV-}: direct \ch{C-F} bond formation by decarboxylation with xenon difluoride via a radical mechanism, and the Balz--Schiemann reaction, which replaces surface amino groups with fluorine. The two routes were compared by infrared, X-ray photoelectron, energy-dispersive X-ray, Raman and photoluminescence spectroscopy. Both gave a high \ch{NV-} fraction, up to $\sim$85\% on average, which to our knowledge is among the highest reported for surface-terminated nanodiamonds of this size and, in particular, for fluorine termination. We attribute the stabilisation to the fluorination treatment, which removes acceptor-like carboxyl groups and forms \ch{C-F} bonds that give the surface a positive electron affinity.
Frequency-domain relaxometry shows that the fluorinated particles retain a long spin--lattice relaxation time, $733\pm56$ and $712\pm20$~\si{\micro\second} for the \ch{XeF2} and Balz--Schiemann routes, several times the values reported for commercial HPHT nanodiamonds, although shorter than the $1173\pm123$~\si{\micro\second} of the as-received material. Fluorination thus activates near-surface \ch{NV-} centres, which dominate relaxometric sensing, at the cost of a moderate shortening of \ensuremath{T_1}.
\end{abstract}

\begin{keyword}
nanodiamond \sep nitrogen vacancy centre \sep fluorination \sep solution \sep surface chemistry
\end{keyword}

\end{frontmatter}

\section{Introduction}

The nitrogen--vacancy (NV) centre in diamond is among the most extensively studied point defects in the solid state. It consists of a substitutional nitrogen atom adjacent to a carbon vacancy; capture of an additional electron yields the negatively charged state, \ch{NV-} ~\cite{Doherty2013,Gali2019}. The \ch{NV-} centre has attracted particular interest as a versatile platform for quantum sensing and other quantum technologies, largely because it possesses long-lived, optically addressable spin states and can therefore operate as a qubit even at room temperature~\cite{Doherty2013,Gruber1997,Jelezko2004}. Its electron spin can be initialised and read out optically by means of optically detected magnetic resonance (ODMR), providing a convenient interface between photons and single spins~\cite{Gruber1997}. These capabilities underpin a broad range of quantum technologies---from nanoscale magnetometry, thermometry, and electrometry to single-molecule sensing and quantum-information processing~\cite{Aslam2023,Du2024,Thalassinos2025}.

All of these functions presuppose that the centre remains in its negative charge state, since \ch{NV^0} lacks the spin-dependent fluorescence contrast on which optical readout relies. In practice, the charge state fluctuates due to the photoionisation of \ch{NV-} to \ch{NV^0} under the very illumination used for spin initialisation and readout~\cite{Doherty2013, Aslam2013}. A stable \ch{NV-} population is therefore a basic prerequisite for reliable operation. {}Equally important is the spin-coherence time, which sets the duration over which the spin retains phase information and thereby bounds the attainable sensitivity of any coherent measurement. Together, charge-state stability and long coherence times determine the performance of \ch{NV-} centres in many applications \cite{Schirhagl2014,Fan2025,Fujisaku2019}.{}

Magnetic fields, temperature and electric fields are most often measured through changes in the ODMR spectrum, with sensitivities that now reach the single-molecule scale~\cite{Du2024,Aslam2023}. The charge state itself can also serve as the sensing signal because the \ch{NV-}/\ch{NV^0} ratio can respond to the local electrostatic and chemical environment \cite{Karaveli2016,Bourgeois2015,Shields2015}, Charge-state-based sensing has been used, for instance, to monitor the local electrochemical potential, with nanodiamond NV ensembles resolving potential swings as small as $\sim$\SI{20}{\milli\volt} \cite{Karaveli2016}. {}Finally, changes in \ensuremath{T_1} report on nearby fluctuating spins, such as paramagnetic free radicals, and it can be detected all-optically with a simple initialise--wait--read fluorescence sequence, with no microwave driving required~\cite{Fan2025,Nie2021,Tetienne2013}, well suited to biosensing. It has enabled the quantification of free radicals generated in chemical reactions and in single living cells and embryos~\cite{Nie2021,PeronaMartinez2020}, and, when combined with responsive surface coatings, even pH measurements over the range 3--7 \cite{Fujisaku2019}.{}

The sensitivity of all of these sensing schemes, however, depends critically on the proximity of the NV centre to the analyte. The dipolar field of a point-like magnetic target decays as $r^{-3}$ with the separation $r$, therefore, the relaxation rate it induces on the NV spin---which scales with the square of the coupling---falls off as $r^{-6}$ \cite{Tetienne2013,Schirhagl2014}. Accurate devices therefore require shallow centres residing within a few nanometres of the surface. Unfortunately, the very susceptibility to external perturbations that makes the NV centre---and any colour centre with comparable properties---such an excellent sensor also renders shallow centres unreliable. Dangling bonds and sp$^2$-like carbon defects at and near the surface constitute an electronic spin bath whose magnetic noise shortens both the coherence time \ensuremath{T_2} and the relaxation time  \ensuremath{T_1} \cite{Myers2014,Romach2015,Sangtawesin2019}; fluctuating surface charges generate electric-field noise that causes additional dephasing \cite{Kim2015}; and electron-accepting surface states pin the Fermi level and induce upward band bending, which depopulates \ch{NV-} in favour of \ch{NV^0} in the near-surface region~\cite{Hauf2011, broadway2018,Shinei2023}. Consequently, \ch{NV-} stability is strongly governed by surface chemistry and crystallographic orientation~\cite{Janitz2022}.

Considerable effort has therefore been devoted to engineering the diamond surface---of bulk crystals and nanodiamonds (NDs) alike---to improve the spin coherence and charge stability of shallow defects while retaining functionality for applications. Diamond surfaces are routinely cleaned in strongly oxidising acid mixtures, which graft oxygen-containing functional groups, most notably carboxyl (\ch{COOH}) groups, that provide convenient moieties for further chemistry but simultaneously reshape the interfacial band structure \cite{Janitz2022,Barzegar2021}. Indeed, the usual carboxyl, hydroxyl, and hydrogen terminations tend to destabilise \ch{NV-} through band bending and unfavourable electron-affinity effects~\cite{Janitz2022,Hauf2011,FreireMoschovitis2023}. Since the as-oxidised surface is thus far from optimal, numerous strategies for its further modification are being pursued.

Among these strategies, nitrogen-terminated surfaces have attracted growing interest owing to their ability to suppress electron spin noise and to promote \ch{NV-} charge-state stability \cite{Janitz2022,Kawai2019,Gaoxian2024,Laube2017,Abendroth2022}. Nitrogen termination has been achieved both by physical routes, such as plasma treatment \cite{Gaoxian2024,Koch2011,Artemenko2019,Jegenyes2025}, and by wet-chemical routes \cite{Laube2017,Sotowa2004,Miller1995,Kaviani2014,Malkinson2024,Jegenyes2025,Czene2023}. Plasma methods, however, can reduce the NV concentration and co-introduce oxygen \cite{Koch2011,Artemenko2019}. Intriguingly, such mixed N/O terminations can themselves further enhance \ch{NV-} stability \cite{Gaoxian2024,Malkinson2024,Rietwyk2013,Czene2025}.

Beyond covalent termination, encapsulation in inorganic shells offers a complementary route: silica coatings and \ch{Al2O3} layers passivate the surface and stabilise the NV charge state~\cite{Barzegar2025, Kumar2024}. Such surface engineering has recently extended the achievable \ensuremath{T_1} of NDs substantially, reaching \SI{4.7}{\milli\second} in \SI{100}{\nano\metre} NDs~\cite{Mameli2026}, $\sim$\SI{2}{\milli\second} after molten \ch{KNO3} oxidation~\cite{Alkahtani2025}, and \SI{1}{\milli\second} in \SIrange{50}{60}{\nano\metre} NDs upon silica coating \cite{Barzegar2025}.

Although many terminations have been examined experimentally and theoretically, fluorine termination is widely regarded as one of the most favourable, because its positive electron affinity (PEA) strongly stabilises the \ch{NV-} charge state~\cite{Janitz2022,Rietwyk2013,Cui2013}. Fluorine plasma treatment increased the stability of shallow ($<\SI{3}{\nano\metre}$) NV centres approximately four-fold, from \SI{0.11}{\percent} to $\approx$\SI{0.45}{\percent} \cite{Osterkamp2013}, and \ch{SF6}-plasma treatment proved essential for stabilising nitrogen-delta-doped NV centres residing within \SI{5}{\nano\metre} of the surface \cite{Osterkamp2015}. Consequently, the development of efficient yet mild fluorination procedures is of considerable interest.

The formation of \ch{C-F} bonds, however, ordinarily demands highly reactive reagents, such as \ch{F2} or \ch{ClF3}, or otherwise harsh conditions, and controlled fluorination of diamond therefore remains a challenge. Owing to its extreme reactivity, fluorination with \ch{F2} requires specialised equipment and careful process control \cite{Greenwood2012,Liu2004,Ando1995}. Even though, \ch{ClF3} provides even stronger fluorinating capability while enabling controlled \ch{C-F} bond formation on diamond surfaces without significant structural degradation \cite{Greenwood2012,Kealey2001}, it is still far from being a safe reaction alternative. To circumvent the handling difficulties of these aggressive reagents, milder electrophilic fluorinating agents have been developed. The most prominent are Selectfluor (F-TEDA-BF$_4$; 1-chloromethyl-4-fluoro-1,4-diazoniabicyclo[2.2.2]octane bis(tetrafluoroborate)), that delivers electrophilic fluorine under mild conditions \cite{Nyffeler2005}, and $N$-fluorobenzenesulfonimide (NFSI), a $N$--F reagent of attenuated reactivity that is valued for selective fluorinations \cite{Differding1991}. In particular, Rodgers \emph{et al.} introduced a gentle, visible-light-driven C--H activation route that fluorinates the diamond surface under ambient conditions while preserving the coherence of NV centres lying within \SI{10}{\nano\metre} of the surface~\cite{Rodgers2024}. Moreover, a carboxyl-selective, silver-catalysed Hunsdiecker--Borodin-type reaction employing Selectfluor was shown to increase the \ch{NV-} fraction of the total NV population in NDs~\cite{Havlik2016}. Recently, combined \ch{Cl2} and \ch{F2} gas treatment of detonation NDs has been used to obtain ultrapure surfaces with fluorine grafted onto the diamond core~\cite{Henry2026}, and for \SI{40}{\nano\metre} fluorescent NDs the immediate surface termination was shown to govern the \ch{NV-} charge stability and \ensuremath{T_1}~\cite{Sadzak2026}.

Despite these efforts, reports in which efficient fluorination is accompanied by a substantial improvement of NV properties remain scarce. For nanodiamonds, solution-phase fluorination has so far afforded only moderate gains in the \ch{NV-} fraction~\cite{Havlik2016}, while the photochemical route of Rodgers \emph{et al.} preserved---but did not extend---the coherence of shallow centres in bulk diamond~\cite{Rodgers2024}. To the best of our knowledge, a mild solution-phase fluorination that markedly improves the \ch{NV-} charge-state stability in nanodiamond has not yet been reported{}, nor has the NV spin--lattice relaxation time of fluorinated nanodiamonds{}.

We previously reported an effective nitrogen termination route of nanodiamonds through the Hofmann degradation, which additionally reduces the sp$^2$ carbon content and thereby affords a significant improvement in charge stability \cite{Czene2023,Jegenyes2025,Czene2025}. While the mixed N/O termination and the diminished graphitic content already improve the NV properties, the amino termination offers a further advantage: it constitutes a convenient functional group for surface fluorination under reaction conditions far milder than those required for oxygen-terminated surfaces. Nevertheless, the development of other fluorination reaction paths remains important.

In this work, we exploit two complementary solution-phase strategies to fluorinate nanodiamond surfaces and compare them directly: the Balz--Schiemann reaction via diazotisation of the surface amino groups and the fluorination with xenon difluoride (\ch{XeF2}).
Xenon difluoride is a particularly attractive fluorinating agent, often showing higher reactivity than Selectfluor or NFSI depending on the reaction conditions \cite{Ramsden2013,ChatalovaSazepin2016,Tramsek2007}. Diamond fluorination with XeF2 is generally attributed to dissociation into highly reactive fluorine-containing species \cite{Foord2001,Morar1986}, whereas reactions in solution frequently proceed via a single-electron transfer mechanism \cite{Ramsden2013,Tius1995,Patrick1993}. Both routes achieve a sub-monolayer fluorine coverage under comparatively mild conditions. We characterise the resulting fluorine-terminated nanodiamonds as a function of fluorine incorporation and examine the stability of the NV charge state in detail, showing that fluorination markedly increases the \ch{NV-} fraction{}, while the spin--lattice relaxation time (\ensuremath{T_1}) is moderately shortened{}.


\section{Methods}
\paragraph{synthesis}
\textbf{The Balz--Schiemann (BS) route.} HPHT diamond particles with a diameter of \SI{140}{\nano\metre} and NV content of \SI{3}{\ppm} were purchased from Adámas Nanotechnologies. According to the manufacturer's dynamic light scattering data, the size distribution is centred at \SI{140}{\nano\metre} and extends from about 100 to \SI{200}{\nano\metre}; the production and properties of these fluorescent diamond particles are reviewed in Ref.~\cite{Shenderova2019}. Amino-terminated diamond was synthesized as follows. From suspension (\SI{1}{\milli\gram\per\milli\litre}), \SI{1}{\milli\litre} was transferred to a flask and evaporated to dry, and \SI{5}{\milli\litre} of \ch{SOCl2} (Sigma Aldrich) was added. The mixture was sonicated in an ultrasonic bath at \SI{37}{\kilo\hertz} for 3 hours at \SI{50}{\celsius}. The solvent was then evaporated under vacuum at \SI{50}{\celsius}. Subsequently, \SI{6}{\milli\litre} of an \ch{NH3}/dioxane mixture (\SI{2}{\milli\litre} of \SI{0.4}{\molar} \ch{NH3}/dioxane diluted in \SI{4}{\milli\litre} of anhydrous dioxane) was added under an argon flow and sonicated under the same conditions. The dioxane mixture was removed under vacuum at \SI{50}{\celsius} and the nanoparticles were redispersed in \SI{6}{\milli\litre} of water, which was followed by sonication at \SI{37}{\kilo\hertz} for 2 hours. The resulting amide intermediate was heated to \SI{80}{\celsius}, and \SI{0.5}{\gram} \ch{NaOH} was added to mixture (VWR) and \SI{0.4}{\milli\litre} \ch{Br2} (Sigma Aldrich) was added dropwise. The solution was stirred at \SI{80}{\celsius} for 15 minutes and then sonicated at \SI{80}{\celsius} for 10 minutes. Excess bromine was neutralised with ammonia solution, and the sample was purified by centrifugation until the pH reached approximately 8. For diazotisation, \SI{200}{\micro\litre} of HF/pyridine (70/30\%, Sigma-Aldrich) was added dropwise to the slightly alkaline suspension in a polypropylene tube and sonicated for 10 minutes at room temperature. Saturated \ch{NaNO2} (VWR) solution was added dropwise, keeping the temperature between \SIrange{0}{2}{\celsius} in an ice bath, with brief ultrasonic exposure for short 10-second periods. Excess \ch{NaNO2} was detected using filter paper soaked in KI/starch. After the mixture reached room temperature, the tube was immersed in boiling water, resulting in a solution temperature of \SIrange{87}{88}{\celsius} for 20 minutes, then it was sonicated at the same temperature for 10 minutes. The cooled mixture was purified by nine cycles of centrifugation, removal of the supernatant and redispersion in ultrapure (Milli-Q) water, each cycle corresponding to an approximately 12-fold dilution. Sample prepared by the described reaction route was labeled from now on as ND\_NH$_2$-F.

\textbf{The XeF2 route.} \SI{1}{\milli\litre} of ND suspension was evaporated to dryness, placed in a Teflon crucible with \SI{100}{\milli\gram} of \ch{XeF2} (Thermo Fisher Scientific), and exposed to an argon stream. Subsequently, \SI{1}{\milli\litre} of \ch{CDCl3} (Sigma Aldrich) and \SI{350}{\micro\litre} of HF/pyridine complex (Sigma Aldrich) were added. Deuterated chloroform was used to avoid the formation of C-H bonds. In the presence of non-deuterated solvents, the possible side-reactions decrease the yield of C-F bonds \cite{Patrick1993,Ramsden2013,ChatalovaSazepin2016}. The crucible was sealed in a bomb tube and heated in a preheated jacket at \SI{100}{\celsius} for 70 hours. The diamond was gradually transferred to the aqueous phase by adding isopropanol/water and re-evaporating in a warm ultrasonic bath. By-products were removed by the same nine centrifugation--redispersion cycles in ultrapure water as for the BS route. For XPS, a portion of ND\_XeF$_2$ was additionally washed with \ch{NaOH}, rinsed with water to neutral pH, treated with \ch{HCl} and washed again by nine centrifugation--redispersion cycles (overall dilution ${\approx}\num{5e9}$).

\paragraph{FTIR} Infrared spectra were recorded in transmission on an evacuated Bruker IFS~66v/S spectrometer equipped with an MCT detector, between 400 and \SI{4000}{\per\centi\metre} at \SI{4}{\per\centi\metre} resolution (256 scans), at room temperature. After purification, \SI{100}{\micro\litre} of a \SI{0.2}{\milli\gram\per\milli\litre} suspension was drop-cast onto a piece of a Czochralski Si wafer (B-doped, (100), \SI{380}{\micro\metre}). Each spectrum was measured against a bare piece of the same wafer as background. The wafer itself absorbs at \SI{611}{\per\centi\metre} (lattice band) and at \SI{1107}{\per\centi\metre} (interstitial oxygen, O$_i$). The strength of the O$_i$ band varies slightly between wafer pieces, so a residue of either sign can remain near \SI{1107}{\per\centi\metre}, and any sample absorption in this region cannot be quantified. Control measurements (empty beam, bare plate against bare plate, and plates dipped into HF/pyridine) are shown in Appendix~\ref{app:ftir}. For Fig.~\ref{fig:ftir}, a rubberband baseline was subtracted from each spectrum, which was then smoothed with a 7-point Savitzky--Golay filter; no normalisation was applied.

\paragraph{XPS} X-ray photoelectron spectroscopy was carried out with an Escalab Xi+ instrument (Thermo Fisher Scientific, Waltham, MA, USA). A microfocused monochromatic Al~K$\alpha$ X-ray gun was used; the diameter of the analysed area was \SI{200}{\micro\metre}. Charge compensation was applied. The base pressure of the analysis chamber was \SI{3e-10}{\milli\bar}. Peak fitting was performed using CasaXPS version 2.3.19 rev1.0m, with Shirley background subtraction and Gaussian--Lorentzian GL(30) peak shapes.

\paragraph{EDS} Fluorine distribution was measured by energy-dispersive X-ray spectroscopy (EDS). The spectra were recorded with a TESCAN MIRA3 electron microscope equipped with an Element EDS system. The applied accelerating voltage was \SI{8}{\kilo\volt}. The relatively low voltage minimised beam damage and was sufficient to excite the K lines of all light elements and of Ca. At this voltage the electron range in diamond (${\approx}\SI{0.6}{\micro\metre}$) exceeds the particle size, so the spectra average over whole particles~\cite{Goldstein2018}. Elemental contents were obtained with the standardless eZAF quantification of the EDAX APEX software. Standardless quantification of light elements is only semi-quantitative, and the EDS data are therefore used qualitatively. EDS was performed on ND\_XeF$_2$, the sample with the higher fluorine content.

\paragraph{PL and Raman} A Renishaw inVia Raman Microscope was used to record photoluminescence (PL) and Raman spectra with a 50× Leica objective (NA\,=\,0.5). The laser power density was calculated as the laser power divided by the area of the Airy disc, of diameter $d = 1.22\lambda/\mathrm{NA} \approx \SI{1.3}{\micro\metre}$ at \SI{532}{\nano\metre}. The excitation wavelength for PL measurements was \SI{532}{\nano\metre} between 535-900 \,\si{\nano\metre}, while Raman spectra were recorded between 1053-2081\,\si{\per\centi\metre} using \num{325} and \SI{785}{\nano\metre} laser sources. The resolution was \,\SI{1}{\per\centi\metre}. These measurements were done on the same samples prepared for FTIR and EDS.

\paragraph{ODMR} For ODMR measurements, the samples were evaporated onto a \SI{150}{\micro\metre}-thick glass plate. CW-ODMR measurements were performed using a home-built optical setup based on a Newport integrating sphere (IS) 819C-IS-5.3 (IS). The sample was placed on a coplanar waveguide antenna and connected to the bottom port of the IS. A \SI{520}{\nano\metre} fibre-coupled laser from Roithner, delivering the excitation to an about \SI{6}{\milli\metre} diameter spot through a \SI{550}{\nano\metre} dichroic mirror (Thorlabs DMLP550T), served as an excitation source and was connected to a side port of the IS. The CW-ODMR spectra were recorded at a laser power of \SI{220.5}{\milli\watt}, corresponding to a power density of approximately \SI{0.78}{\watt\per\centi\metre\squared} assuming uniform illumination. The radiation from the sample was collected through a biased silicon-based free-space photodetector (Thorlabs DET100A2) using a \SI{700}{\nano\metre} longpass hard-coated filter (Thorlabs FELH0700) and a \SI{645}{\nano\metre} longpass coloured-glass filter (Thorlabs FGL645M), and connected to the top port of the IS. The photocurrent from the detector output was converted into voltage using a \SI{1}{\mega\hertz} bandwidth transimpedance amplifier (TIA) (Femto DHPCA-100) and connected to the input of a lock-in amplifier (Anfatec USBLockIn250). The microwave (MW) field near the ground-state spin resonance frequency ($2870\pm70$~\si{\mega\hertz}) was generated by a MW generator (Vaunix LabBrick LSG-402) in conjunction with a high-power amplifier (Mini-Circuits ZHL-25W-63+) and applied to the sample through a coplanar waveguide terminated with a high-power (\SI{100}{\watt}) \SI{50}{\ohm} termination. The microwave excitation amplitude was modulated using a high-speed switch (Mini-Circuits ZASWA-2-50DRA+) and controlled by a TTL signal from the internal reference source of the lock-in amplifier. The microwave power applied for the CW-ODMR spectra was \SI{-36}{\dBm} (\SI{251}{\nano\watt}).
{}The spin--lattice relaxation time \ensuremath{T_1} was determined with the same CW-ODMR setup by frequency-domain relaxometry, following the protocol of Verkhovlyuk \emph{et al.}~\cite{Verkhovlyuk2026}. Instead of a pulsed initialise--wait--read sequence, the amplitude of the microwave field was modulated and the lock-in response was recorded while sweeping the modulation frequency from about \SI{10}{\hertz} to \SI{20}{\kilo\hertz}. The response passes through a maximum at a characteristic modulation frequency $f_{\mathrm{max}}$ that is governed by the relaxation dynamics of the spin ensemble. The microwave generator was set to \SI{-20}{\dBm}, corresponding to \SI{25}{\dBm} at the sample after the amplifier, and $f_{\mathrm{max}}$ was determined at four laser powers spanning \SIrange{142}{354}{\milli\watt} (approximately \SIrange{0.50}{1.25}{\watt\per\centi\metre\squared}). Identical microwave conditions were used for all three samples. Because continuous optical illumination itself contributes to the measured relaxation, $f_{\mathrm{max}}$ was plotted as a function of laser power and fitted with a straight line; the intercept $f_{0}$ of this extrapolation to zero laser power yields the intrinsic relaxation time as $T_1 = 1/(2\pi f_{0})$. All \ensuremath{T_1} values quoted below are such zero-power extrapolations. The approach retrieves \ensuremath{T_1} over more than three orders of magnitude and remains applicable to inhomogeneous ensembles such as nanodiamond powders, while offering a substantial speed-up over pulsed relaxometry~\cite{Verkhovlyuk2026}.{}

\paragraph{Data analysis} The raw data were analysed and plotted with Python~3 (NumPy, SciPy, Matplotlib). The scripts were written with the assistance of the generative AI model Claude (Anthropic) and are available from the corresponding author.

\section{Results and discussion}
\subsection{Synthesis and chemical characterization} 
\begin{figure}[htbp]
      \centering
      \includegraphics[width=\linewidth]{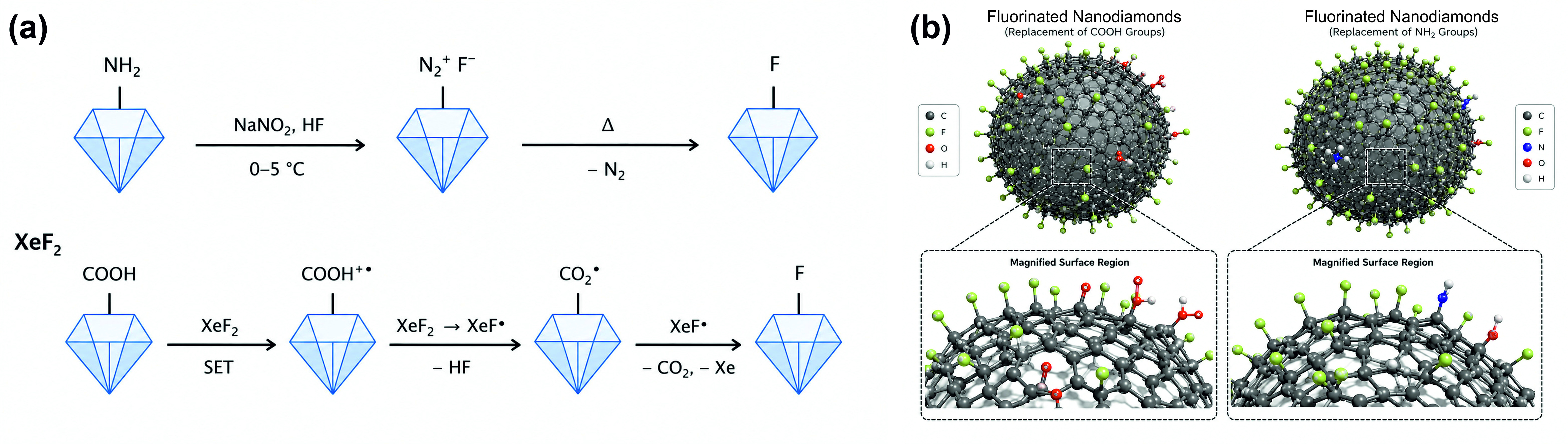}
      \caption{(a) The fluorination reactions: the Balz--Schiemann (BS) reaction and the \ch{XeF2}-based route. The reactions show the primary attachment sites; in the \ch{XeF2} route, fluorine also attaches at other sites (Section~3.1). (b) Simplified overview of the surface-modified diamond. The real surface is heterogeneous: fluorine covers well below one monolayer, alongside residual oxygen-containing groups and sp$^2$-carbon patches (Section~3.1).}
      \label{fig:reactions}
    \end{figure}

The processes of the two reactions are shown in Fig.~\ref{fig:reactions}, which also schematically presents the surface.

\begin{figure}[htbp]
      \centering
      \includegraphics[width=0.75\linewidth]{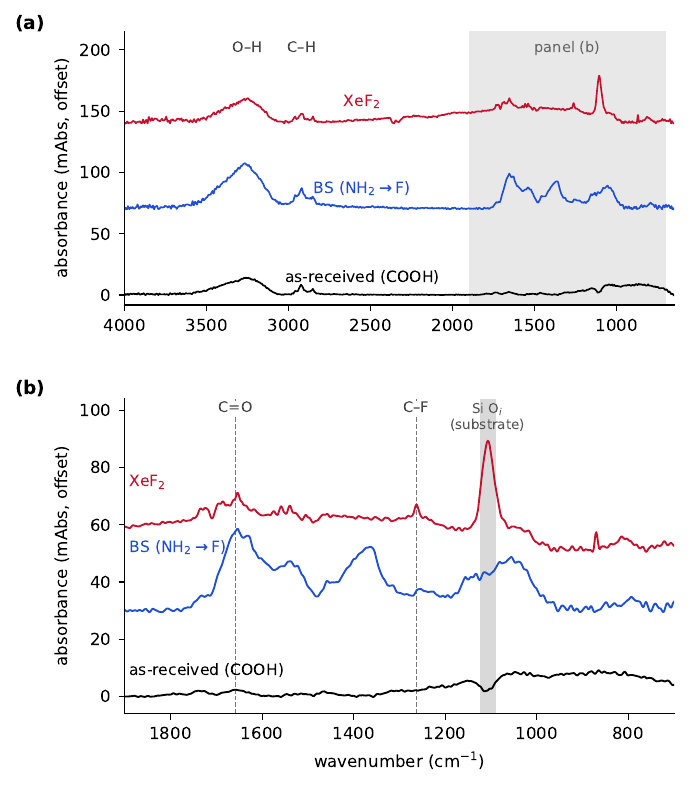}
      \caption{FTIR spectra of the as-received (\ch{COOH}-terminated) and the two fluorinated nanodiamonds, recorded under vacuum against a bare Si plate. (a) Full range; (b) the \SIrange{700}{1900}{\per\centi\metre} region. The spectra are shown on a common absolute scale and offset for clarity. The shaded band marks the interstitial-oxygen band of the Si substrate at \SI{1107}{\per\centi\metre}, whose residue varies between wafer pieces; the sample contribution in this region cannot be quantified. The sharp feature at \SI{870}{\per\centi\metre} in the \ch{XeF2} spectrum is instrumental.}
      \label{fig:ftir}
    \end{figure}

\paragraph{FTIR} The FTIR spectra of the three samples are shown in Fig.~\ref{fig:ftir}. After fluorination, a new, sharp band appears at \SI{1263}{\per\centi\metre}. It is strongest in ND\_XeF$_2$, weak in ND\_NH$_2$-F, and absent both in the as-received material and on bare or HF/pyridine-treated Si plates (Appendix~\ref{app:ftir}). Its intensity thus follows the fluorine content determined by XPS (Table~\ref{tab:xps}). The films differ in thickness, but the order is unchanged when the band heights are referred to the amount of material on each plate, estimated from the scattering background of the film; the intensity ratio, however, is not quantitative. We tentatively assign it to the stretching vibration of covalent \ch{C-F} bonds on sp$^3$ carbon. Fluorographene prepared with the same reagent, \ch{XeF2}, shows its characteristic covalent \ch{C-F} band at \SI{1260}{\per\centi\metre}~\cite{Jeon2011}, and molecular-dynamics simulations place the \ch{C-F} stretching mode of fluorinated sp$^3$ carbon at \SIrange{1200}{1350}{\per\centi\metre}~\cite{Ueta2012c}. On F$_2$-fluorinated diamond and nanodiamond, \ch{C-F} bands have been reported at 1251--\SI{1258}{\per\centi\metre}~\cite{Ando1995,Osipov2020} and in the range 1200--\SI{1260}{\per\centi\metre}~\cite{Zagrebina2015}. The dominant band near \SI{1340}{\per\centi\metre} of extensively F$_2$-fluorinated nanodiamond~\cite{Liu2004,Osipov2020} is not observed, consistent with the sub-monolayer fluorine coverage of the present samples.

\ch{C-F} bonds on fluorinated nanodiamond are also reported near \SI{1100}{\per\centi\metre}~\cite{Liu2004,Zagrebina2015}. In our spectra the band near \SI{1100}{\per\centi\metre} probably contains a contribution from the samples, superimposed on the interstitial-oxygen band of the Si substrate (shaded in Fig.~\ref{fig:ftir}b). Because the substrate residue varies between wafer pieces, the sample contribution cannot be quantified, and we do not use this band as primary evidence. A weak feature near \SI{800}{\per\centi\metre} is present in the as-received material as well and is therefore not related to fluorination. No significant band appears above \SI{1800}{\per\centi\metre}, where carbonyl fluoride (C(O)F) groups would absorb. The band at \SI{1658}{\per\centi\metre} is assigned to C=O stretching of residual carbonyl groups, with a possible contribution of adsorbed water~\cite{Petit2018,Coates2006}. ND\_NH$_2$-F shows an additional broad band near \SI{1365}{\per\centi\metre}, absent in the other samples, which we cannot assign unambiguously; it may arise from nitrogen-containing groups remaining from the amination route or from the carbonaceous by-product of this route. The band assignments are summarised in Table~\ref{tab:ftir} (Appendix~\ref{app:ftir}). The identification of \ch{C-F} bonds therefore rests primarily on the XPS spectra discussed below, and the FTIR results are consistent with it.

In aqueous media, the fluoride ion is strongly hydrated; therefore, HF behaves as a weak nucleophile only \cite{Nolte2012}, so it is not suitable for effective replacement of surface OH groups. Previous studies have shown that anhydrous HF alone does not result in significant fluorination on the diamond surface \cite{Kealey2001}. Accordingly, the role of the pyridine--HF system employed in the present work is not to serve as a fluorine source but rather to act as a catalyst and a solvent for \ch{XeF2} \cite{Ramsden2013,Tramsek2007}.

\begin{figure}[htbp]
      \centering
      \includegraphics[width=0.7\linewidth]{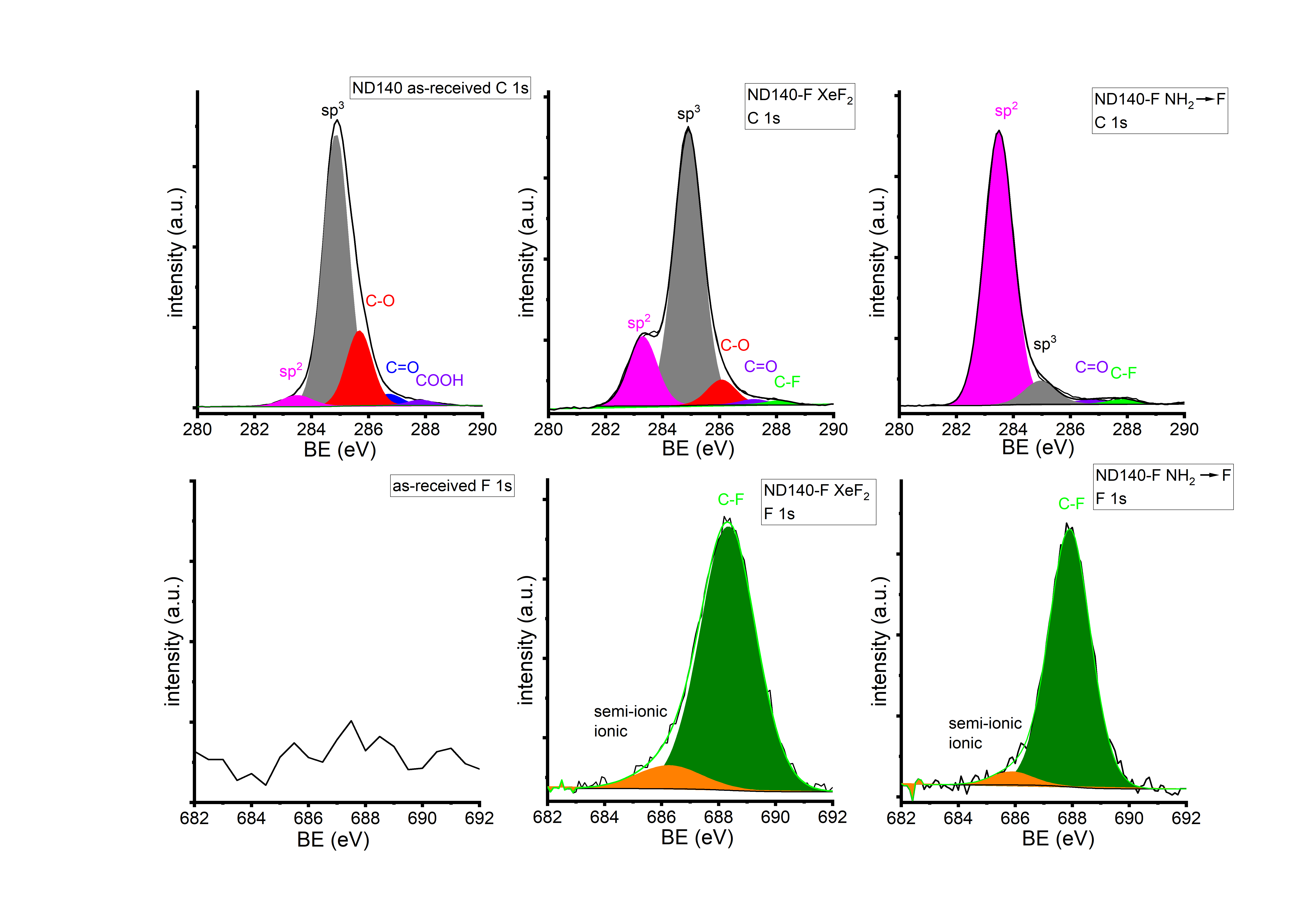}
      \caption{XPS spectra of the two fluorinated and the as-received \ch{COOH}-terminated samples.}
      \label{fig:xps}
    \end{figure}

\paragraph{XPS} The high-resolution C\,1s and F\,1s XPS spectra (Fig.~\ref{fig:xps}) also indicate that the applied reactions created \ch{C-F} bonds on the nanodiamond surface. The C\,1s spectrum of the as-received sample was dominated by the sp$^3$ carbon peak at around \SI{285}{\electronvolt}, accompanied by a small sp$^2$ contribution at \SI{283.5}{\electronvolt}, i.e.\ at lower binding energy than the sp$^3$ line, as commonly observed for nanodiamond \cite{Romanyuk2024,Panich2010}. The as-received sample also exhibited \ch{C-O} and \ch{C=O} components. After fluorination, the C\,1s spectra contain a component near \SI{288.2}{\electronvolt} to which the \ch{C-F} carbon contributes~\cite{Foord2001,Hantsche1993,Panich2010}. This component overlaps with carbonyl carbon (\SI{288.0}{\electronvolt} in the as-received material) and amounts to only ${\approx}\SI{1}{\percent}$ of the C\,1s signal, so the bonding of fluorine is identified from the F\,1s spectra discussed below. The most pronounced change was the growth of the sp$^2$ component. In the analysis spots shown in Fig.~\ref{fig:xps}, the sp$^3$:sp$^2$ peak-area ratio decreased from ${\approx}36$ in the as-received material to ${\approx}4$ in ND\_XeF$_2$ (${\approx}2$ in a second analysis spot) and to ${\approx}0.1$ in ND\_NH$_2$-F. We attribute the high sp$^2$ content of ND\_NH$_2$-F to a carbonaceous by-product that could not be removed by the applied purification protocol. This by-product is largely nitrogen-free: with \SI{0.3}{at.\percent} N, adsorbed pyridine could account for at most ${\approx}\SI{1.5}{at.\percent}$ of the carbon. ND\_NH$_2$-F was washed repeatedly with water---each cycle consisting of centrifugation, removal of the supernatant and redispersion, which corresponds to an approximately 12-fold dilution---and monitored by XPS after each washing cycle; no significant changes were observed in the monitored spectral windows. The presence and significance of this sp$^2$ carbon will be discussed below. Here we merely note that the C\,1s components appear at the same positions in all samples, with varying relative weights.
In the F\,1s spectra, both samples exhibited the same two components, at \SI{688.2}{\electronvolt} and \SI{686}{\electronvolt}, binding energies characteristic of covalent and semi-ionic \ch{C-F} bonding environments, respectively~\cite{Foord2001,Hantsche1993,Kang2018}. The covalent component dominates (\SI{91}{\percent} for ND\_XeF$_2$ and \SI{95}{\percent} for ND\_NH$_2$-F in the spectra of Fig.~\ref{fig:xps}; the semi-ionic share of ND\_XeF$_2$ varied between 9 and \SI{17}{\percent} across analysis spots). The Si\,2p spectra contain no component in the range of silicon fluorides or fluorosilicates ($\gtrsim$\SI{104.5}{\electronvolt}), so the \SI{686}{\electronvolt} component is not due to fluorinated substrate silicon. The quantity of the measured elements is indicated in Table~\ref{tab:xps}.

It should also be mentioned that the absolute positions of the recorded photoelectron peaks cannot be determined free of doubt: highly insulating diamond particles are subject to charging and, in particular, to non-uniform (differential) charging, even when charge compensation with a low-energy electron flood gun is applied \cite{Baer2020,Greczynski2020}. All spectra were recorded with charge compensation and evaluated without any subsequent binding-energy correction; in the high-resolution C\,1s spectra the diamond sp$^3$ component appeared at ${\approx}\SI{285}{\electronvolt}$ in all samples. The survey spectra, recorded with a larger pass energy at the end of each measurement series, were not used for energy referencing. Minor low-binding-energy components common to several core levels are interpreted with caution. The low-binding-energy C\,1s component of ND\_XeF$_2$ is, however, not a charging artefact: it persisted under varied charge-compensation settings.

    \begin{table}[htbp]
      \centering
      \caption{F, C, O, Si, N, Cl and Br content (at.\%) determined by XPS for the as-received sample and for the F-terminated samples prepared by the \ch{XeF2} and BS methods. The \ch{XeF2} values are the average of two analysis spots on the portion additionally washed with \ch{NaOH} and \ch{HCl} (Section~2); N varied between the spots from the detection limit to ${\approx}\SI{2}{at.\percent}$. The as-prepared material gave similar values. Cl in the \ch{XeF2} sample most likely originates from the \ch{CDCl3} solvent, Br in the BS sample from the Hofmann degradation (\ch{Br2}/\ch{NaOH}).}
      \label{tab:xps}
      \begin{tabular}{ c c c c c c c c c}
        \toprule
        Sample & C (at.\%) & F (at.\%) & O (at.\%) & Si (at.\%) & N (at.\%) & Cl (at.\%) & Br (at.\%) & F:C ratio\\
        \midrule
        \ch{COOH}          & 91.6 & 0 & 8.1 & 0.3 & 0 & 0 & 0 & 0\\
        \ch{XeF2}          & 66.3 & 2.6 & 16 & 15 & 1 & 0.8 & 0 & 3.9\\
        \ch{NH2 -> F} (BS) & 79.8 & 0.7 & 12 & 6.4 & 0.3 & 0 & 1.1 & 0.9\\
        \bottomrule
      \end{tabular}
    \end{table}

According to XPS, ND\_XeF$_2$ incorporated markedly more fluorine than ND\_NH$_2$-F, with measured F concentrations of \SI{2.6}{\percent} and \SI{0.7}{\percent}, respectively. The lower F content in the BS reaction can be explained by several factors. The main reason, which we have already seen in our earlier experiments, is that the Hofmann degradation is not complete: unlike SiC, there are significantly more unreacted \ch{COOH} groups according to the FTIR \cite{Jegenyes2025,Czene2023}. Another important factor is the stability of the diazonium salt. Halliwell and Nyburg pointed out that the counterion influences its stability \cite{Halliwell1960}. For example, diazonium salts containing smaller anions are more susceptible to hydrolysis, resulting in the formation of phenols rather than the corresponding haloarenes. The decrease in the carbon content can be attributed partly to surface reactions with carbon atoms and partly to the reduced coverage of the silicon substrate. The reduced coverage explains the higher Si content. This difference is clearly evident in the XPS C 1s spectra, whereas it can only be inferred from the FTIR data. Both fluorinated products exhibit a substantial increase in the proportion of \ch{sp^2}-hybridized carbon. The lower carbon content can be attributed to the lower coverage of the Si plate. It should be noted, however, that XPS is highly surface-sensitive, and the measured composition may therefore overrepresent surface-enriched species. More than two-thirds of the signal originates from the top \SI{3}{\nano\metre} depth layer, while nearly 80\% comes from the top \SI{5}{\nano\metre}. The observed increase in \ch{sp^2} carbon is likely more dominant at the surface and may not be fully representative of the complete crystal \cite{Carlson1972,Powell2020}. Removing a single-nm layer can result in a signal reduction of up to 50\% \cite{Baer2005}. Ducrozet and co-workers aimed to coat a diamond with an \ch{sp^2} C layer like an onion skin; although the XPS signal was significant, the actual \ch{sp^2} shell was only a single layer \cite{Ducrozet2021}. The corresponding surface coverages are derived at the end of this section.

\paragraph{Raman} 

\begin{figure}[htbp]
      \centering
      \includegraphics[width=0.8\linewidth]{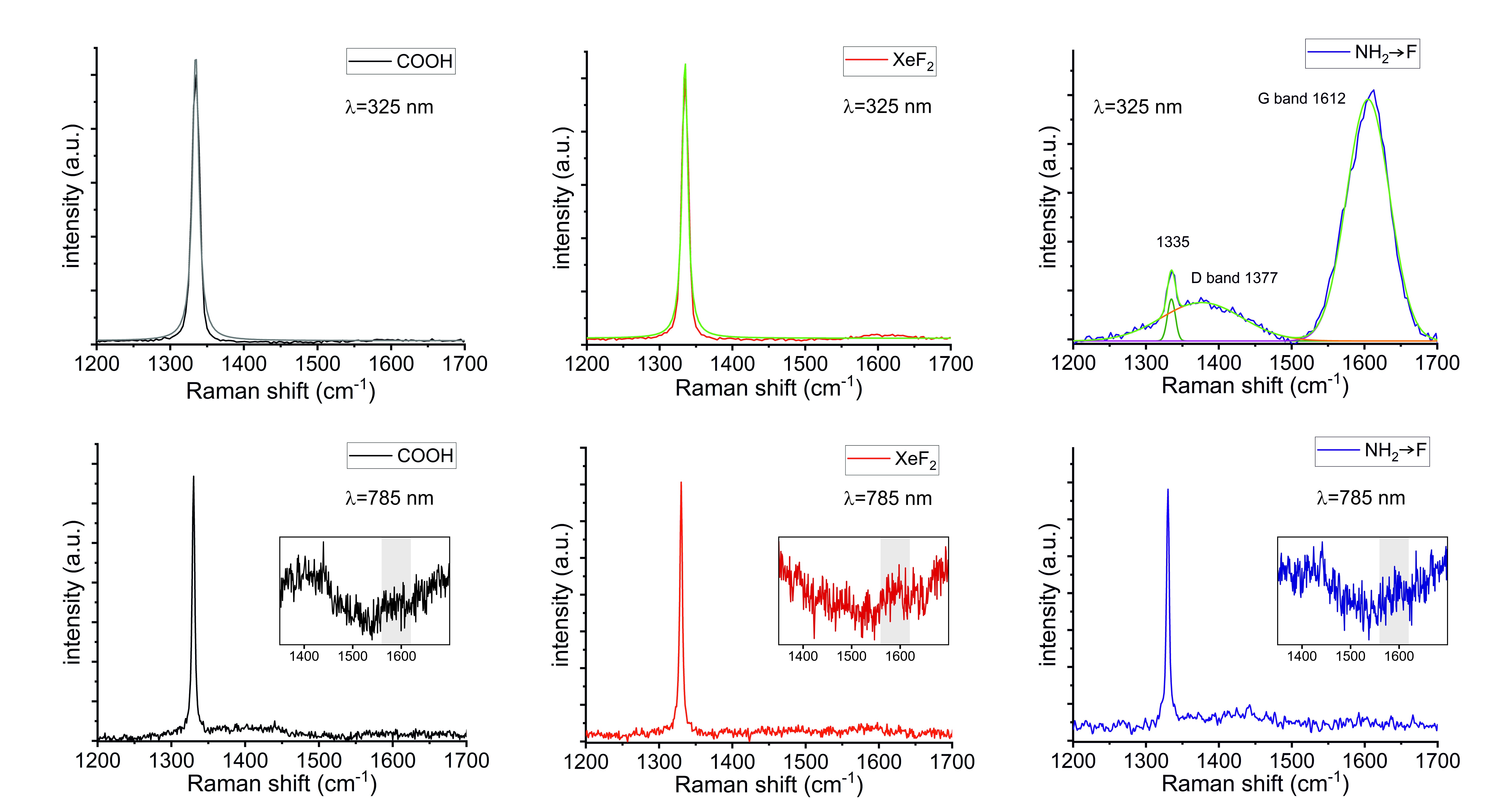}
      \caption{Raman spectra of the \ch{COOH}-terminated sample and the two products after fluorination. The spectra in the top row were measured with a \SI{325}{\nano\metre} laser, the bottom row with a \SI{785}{\nano\metre} laser. Insets: the \SIrange{1350}{1700}{\per\centi\metre} region of the \SI{785}{\nano\metre} spectra, enlarged and baseline-corrected; the grey band marks the position of the G band.}
      \label{fig:raman}
    \end{figure}

The mild nature of the fluorination procedures and the preservation of the diamond structure were verified by Raman spectroscopy. This is a powerful method for investigating the allotropic forms of carbon due to its high sensitivity to the hybridization state of carbon atoms. Furthermore, the use of excitation lasers with different wavelengths allows discrimination between amorphous carbon that is structurally associated with the diamond lattice and amorphous carbon present only at the particle surface \cite{Veres2007}. The Raman spectra are shown in Fig.~\ref{fig:raman}. The characteristic diamond peak is visible in all samples at \SI{1335.5}{\per\centi\metre}, and the G band is at \SI{1613}{\per\centi\metre}. We applied a UV laser because it clearly shows the diamond peak and it is possible to determine the integrity of the diamond lattice. Furthermore, the G band is visible in the UV Raman spectra as well. The latter can easily shift as a result of doping or strain \cite{Ferrari2001}.

Following the \ch{XeF2} treatment, only a slight increase in the intensity of the G band is observed. In contrast, the sample prepared via the BS reaction exhibits a much more pronounced increase. However, the strong enhancement should not be interpreted as a direct measure of the amorphous carbon content. To test this, the Raman spectra were also recorded with a \SI{785}{\nano\metre} laser. Under near-infrared excitation, the BS sample shows only a weak, broad sp$^2$-related feature, much weaker than under UV excitation. This excitation dependence is the opposite of what extended graphitic carbon would produce: at visible and near-infrared excitation, the Raman cross-section of sp$^2$ carbon exceeds that of the sp$^3$ phase by up to two orders of magnitude \cite{FerrariRobertson2000,FerrariRobertson2004}. The weakness of the sp$^2$ bands at \SI{785}{\nano\metre} therefore rules out any significant amount of extended, $\pi$-conjugated graphitic carbon. Their strength at \SI{325}{\nano\metre} (\SI{3.81}{\electronvolt}) shows instead that the sp$^2$ phase consists of small clusters with wide optical gaps of ${\sim}3$--\SI{4}{\electronvolt}, resonantly probed in the UV but far from resonance in the near-infrared \cite{FerrariRobertson2004}. The position of the D-like band supports this. The D band of graphitic carbon shifts by ${\sim}\SI{50}{\per\centi\metre}$ per eV of excitation energy \cite{Matthews1999}, which would place it at ${\sim}1410$--\SI{1425}{\per\centi\metre} at \SI{3.81}{\electronvolt}. The observed \SI{1377}{\per\centi\metre} indicates a strongly suppressed dispersion, characteristic of the ring-breathing modes of small aromatic units \cite{Castiglioni2001a,Castiglioni2001b,FerrariRobertson2000}. Adsorbed aromatic species may also contribute to the UV signal through resonance enhancement \cite{Hughes1949,Wang2009,Veres2007,Rozsa2026}.

Table~\ref{tab:raman} shows the half-width values of the diamond peaks determined with the UV laser. The position (1335.3--\SI{1335.5}{\per\centi\metre}) and width (9.6--\SI{10.8}{\per\centi\metre}) of the first-order diamond line are unchanged within the experimental scatter by either treatment, which shows that the crystalline core is preserved. For particles of this size the diamond Raman line is a more sensitive probe of crystal quality than X-ray diffraction, which averages over the whole particle volume and would not resolve changes confined to the outermost atomic layer.

    \begin{table}[htbp]
      \centering
      \caption{Measured positions and full widths at half maxima of the diamond band ($\lambda = \SI{325}{\nano\metre}$).}
      \label{tab:raman}
      \begin{tabular}{lcc}
        \toprule
        Sample & diamond band (\si{\per\centi\metre}) & half-width (\si{\per\centi\metre}) \\
        \midrule
        \ch{COOH}          & 1335.5 & 10.4 \\
        \ch{XeF2}          & 1335.5 & 9.6  \\
        \ch{NH2 -> F}      & 1335.3 & 10.8 \\
        \bottomrule
      \end{tabular}
    \end{table}

The sp$^2$ carbon in ND\_NH$_2$-F is thus best described as small aromatic islands rather than graphitic shells. The persistence of a prominent diamond line implies a thin, partial surface coverage on intact crystalline cores, in consistence with the sp$^2$ dominated C\,1s XPS spectrum and the absence of related vibration modes in the FTIR data. As a result of the Raman measurements, we could confirm the mild fluorination.

\begin{figure}[htbp]
      \centering
      \includegraphics[width=0.85\linewidth]{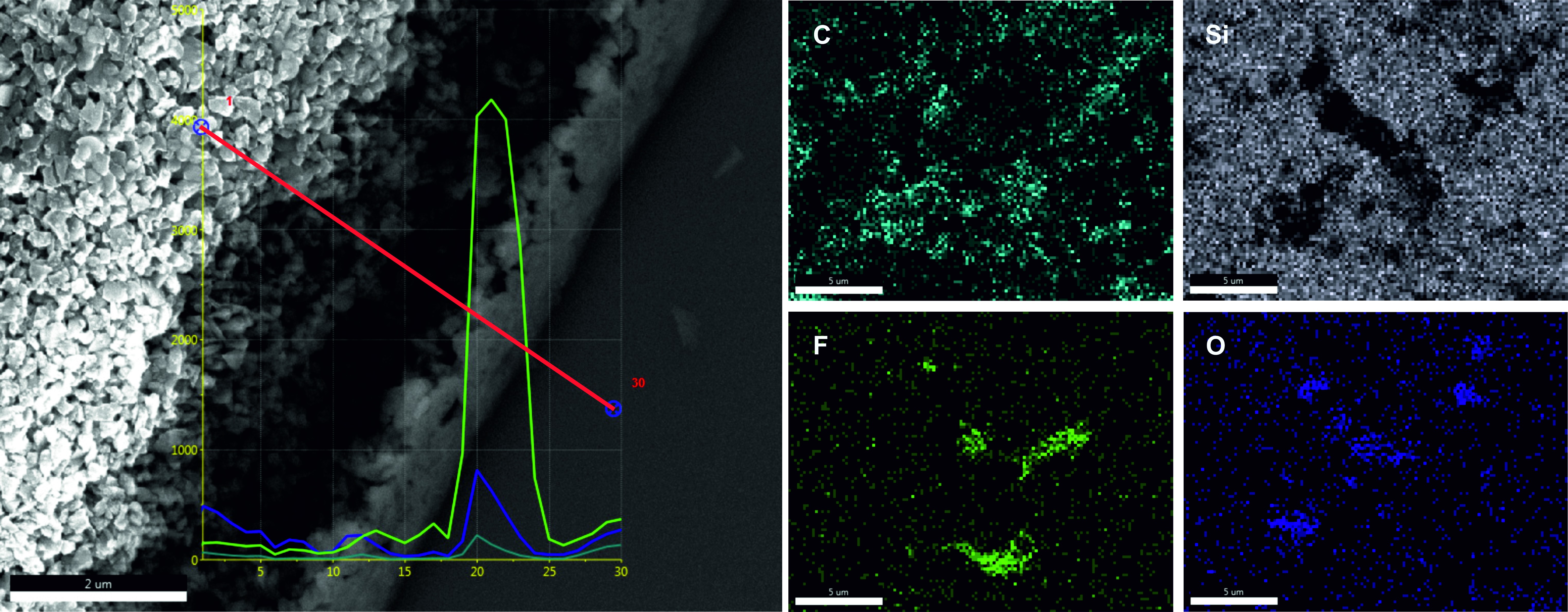}
      \caption{EDS analysis of ND\_XeF$_2$. Left: SEM image at the rim of the drop-cast film with the line-profile scan taken along the red line (30 points). Right: elemental maps of C (turquoise), O (blue), F (green) and Si (grey) recorded in the interior of the film. The fluorine-rich features at the rim contain metal-fluoride residues (see text).}
      \label{fig:eds}
    \end{figure}

\paragraph{EDS analysis} SEM-EDS was used to map the lateral distribution of fluorine across drop-cast films of ND\_XeF$_2$. The two techniques differ in their sampling: XPS averages over a spot of ${\sim}\SI{200}{\micro\metre}$ diameter and over only the outermost few nanometres, whereas at \SI{8}{\kilo\volt} EDS probes the whole volume of the \SI{140}{\nano\metre} particles~\cite{Goldstein2018}. For a particle of radius $r$, a fluorine monolayer corresponds to a volume-averaged ratio F/C\,$\simeq 3\sigma/(n_{\mathrm{C}} r) \approx \SI{0.4}{\percent}$ for spherical particles with $r = \SI{70}{\nano\metre}$ ($\sigma$ and $n_{\mathrm{C}}$ as in Eq.~\eqref{eq:coverage} below). The irregular shape of milled HPHT particles raises the surface-to-volume ratio, but by a factor of order unity~\cite{Havlik2016}, so the coverage of 0.4--0.6 of a monolayer derived from XPS corresponds to F/C of a few tenths of a percent, close to the EDS detection limit for fluorine. Fluorine bonded to the nanodiamond surface is therefore not resolved in the maps, and EDS is used here only to locate fluorine-containing material. The fluorine signal is concentrated in a few micrometre-sized features. At the rim of the drop-cast film, local fluorine contents of 1.3--\SI{4.6}{at.\percent} (F/C\,=\,1.5--\SI{6}{\percent}) were measured. They are accompanied by Ca (1.4--\SI{3.4}{at.\percent}), Mg and Na in amounts sufficient to bind the fluorine as metal fluorides; such salts accumulate at the contact line of a drying drop~\cite{Deegan1997}. The lowest of these values is compatible with fluorinated nanodiamond within the uncertainty of the estimate above, whereas the higher ones require such residues. In the interior, fluorine is concentrated in patches of a few micrometres (Fig.~\ref{fig:eds}), whose local fluorine content exceeds what a monolayer on \SI{140}{\nano\metre} particles can produce by more than an order of magnitude. These patches may contain finer particles with a larger surface-to-volume ratio or a fluorine-rich non-diamond phase, which EDS cannot distinguish. Metal-fluoride residues do not contribute significantly to the XPS fluorine signal: the XPS survey spectra show no Ca\,2p signal (Ca $\leq$\,\SI{0.1}{at.\percent} for ND\_XeF$_2$ and $\leq$\,\SI{0.2}{at.\percent} for ND\_NH$_2$-F), and the F\,1s spectra are dominated by the component at \SI{688.2}{\electronvolt}, whereas Na, Ca and Mg fluorides appear at ${\approx}684.5$--\SI{685.7}{\electronvolt}~\cite{SanzMatias2022}. A metal-fluoride contribution could therefore only be contained in the minor component near \SI{686}{\electronvolt}, which we attribute predominantly to semi-ionic \ch{C-F} (see below). The fluorine contents and coverages discussed below are therefore those obtained by XPS.

\paragraph{Location and bonding of the fluorine} The XPS fluorine contents are atomic fractions averaged over the whole sampling depth, not surface coverages. Fluorine resides in the outermost atomic layer, whereas the C\,1s signal originates from a depth of the order of the inelastic mean free path $\lambda$. For a monolayer of surface density $\sigma$ on a spherical particle much larger than $\lambda$, the expected ratio is
\begin{equation}
  \left.\frac{\mathrm{F}}{\mathrm{C}}\right|_{1\,\mathrm{ML}} \simeq \frac{2\sigma}{n_{\mathrm{C}}\lambda},
  \label{eq:coverage}
\end{equation}
where $n_{\mathrm{C}} = \SI{1.76e23}{\per\centi\metre\cubed}$ is the atomic density of diamond. The factor of two arises because the whole facing hemisphere contributes surface atoms, while the carbon signal is limited to a depth $\lambda$ along the line of sight. With $\sigma \approx \SI{1.8e15}{\per\centi\metre\squared}$ and $\lambda = 2$--\SI{3}{\nano\metre}~\cite{Powell2020}, a full monolayer corresponds to F/C\,$\approx$\,7--\SI{10}{\percent}. The measured F/C ratios of \SI{3.9}{\percent} (ND\_XeF$_2$) and \SI{0.9}{\percent} (ND\_NH$_2$-F) therefore correspond to ${\approx}0.4$--0.6 and ${\approx}0.1$ of a monolayer, respectively. The latter is a lower bound, because the sp$^2$-rich overlayer of ND\_NH$_2$-F attenuates the signal of the diamond surface.

The primary attachment sites are set by the chemistry of the two routes. In the \ch{XeF2} route, fluorine replaces carboxyl groups by radical decarboxylation; in the BS route, it replaces amino groups that were themselves formed from carboxyl groups by the Hofmann degradation. In both cases the \ch{C-F} bond is formed on the sp$^3$ surface carbon that carried the original group. The carboxyl budget of the starting material, however, shows that this cannot be the only pathway for \ch{XeF2}. The C\,1s spectrum of the as-received material contains no resolved O--C=O component; even counting the whole carbonyl component at \SI{288.0}{\electronvolt}, carboxyl-type carbon amounts to at most ${\approx}\SI{1.9}{\percent}$ of the C\,1s signal, i.e.\ to at most ${\approx}0.2$--0.3 of a monolayer according to Eq.~\eqref{eq:coverage}. The fluorine coverage of ND\_XeF$_2$ is about twice as large. Fluorine must therefore also attach at other sites, most plausibly by substitution of hydroxyl groups, which is the main fluorination pathway of nanodiamond with elemental F$_2$~\cite{Zagrebina2015}, and possibly at C--H and sp$^2$ sites. For ND\_NH$_2$-F, the much lower coverage is compatible with a partial conversion of the carboxyl-derived amino groups alone.

The fluorine is present as isolated monofluoride groups. The C\,1s spectra contain no \ch{CF2} component (${\approx}\SI{291}{\electronvolt}$) above \SIrange{0.1}{0.2}{\percent} of the signal in either sample, nor a \ch{CF3} component (${\approx}\SI{293}{\electronvolt}$) in ND\_NH$_2$-F, as expected at sub-monolayer coverage. The dominant F\,1s component at \SI{688}{\electronvolt} corresponds to covalent \ch{C-F}, and the minor one at \SI{686}{\electronvolt} (5--\SI{17}{\percent}) predominantly to semi-ionic \ch{C-F}, with a possible minor contribution from the metal-fluoride residues identified by EDS, which appear in the same energy range~\cite{SanzMatias2022}. In fluorinated carbons, this term denotes \ch{C-F} bonds weakened by interaction with neighbouring unfluorinated carbon, which appear at lower F\,1s binding energy and lower stretching frequency than covalent \ch{C-F}, typically at low fluorine content~\cite{Chronopoulos2017}. At the coverages found here, most CF groups have unfluorinated neighbours (C--H, C--OH or sp$^2$ patches), which accounts for the semi-ionic fraction. The comparison with nanodiamond fluorinated by elemental F$_2$ supports this picture. Low-temperature F$_2$ treatment of detonation nanodiamond produces \ch{CF2}/\ch{CF3} groups on the sp$^2$ shell, with sharp IR bands at 1220 and \SI{1244}{\per\centi\metre}~\cite{Henry2026}; neither these bands nor \ch{CF2} carbon are observed here. F$_2$ at \SI{500}{\celsius}, in contrast, saturates the surface with monofluoride groups bonded to sp$^3$ carbon, with \ch{C-F} bands at 1200--\SI{1260}{\per\centi\metre}~\cite{Zagrebina2015}, the range in which our band at \SI{1263}{\per\centi\metre} lies. Fluorine is therefore bonded predominantly as covalent monofluoride on the sp$^3$ diamond surface, with a minor semi-ionic fraction attributed to isolated CF groups with unfluorinated neighbours and, possibly, to fluorine on sp$^2$ patches.

\subsection{Optical properties}

\paragraph{Origin of the \ch{NV-} stabilisation} Fluorine is strongly electronegative, and the \ch{C-F} bond carries a partial charge from carbon to fluorine. Why fluorination should stabilise the \emph{negative} charge state of NV centres therefore needs an explicit explanation. The partial charge is a polarisation within the \ch{C-F} $\sigma$ bond; it removes no electrons from the diamond bands and creates no free holes. Its macroscopic effect is a layer of surface dipoles with the negative end outward, which shifts the vacuum level relative to the diamond bands and gives the fluorinated surface a positive electron affinity and a large ionisation energy~\cite{Cui2013,Rietwyk2013}. The charge state of an NV centre below the surface, in turn, is set by the position of the Fermi level relative to the NV$(-/0)$ level, which near the surface is governed by band bending~\cite{Hauf2011,broadway2018}. Upward band bending, which converts \ch{NV-} into \ch{NV^0}, has two main causes: surface transfer doping into an adsorbed water layer, which is possible only for surfaces with a small ionisation energy such as hydrogen-terminated diamond, and acceptor-like surface states of carbonyl and carboxyl groups, sp$^2$ defects and dangling bonds, which trap electrons from the nitrogen donors. Fluorine termination suppresses the first mechanism by placing the valence-band maximum far below the redox level of adsorbed water, and weakens the second by replacing \ch{COOH} groups with saturated \ch{C-F} bonds that introduce no states in the gap. The electronegativity of fluorine is therefore the cause of the \ch{NV-} stabilisation, not a contradiction to it. The effect is indirect: fluorination modifies the surface electrostatics, not the NV centres themselves (see the ODMR results below).

\paragraph{PL} Despite the lower fluorine content and the sp$^2$ carbon content of ND\_NH$_2$-F, both fluorination routes produced a substantial increase in the measured \ch{NV-} fraction and its photostability. The \ch{NV^0}/\ch{NV-} ratio was determined from photoluminescence (PL) spectra recorded at laser power densities spanning two orders of magnitude under 532 nm excitation.

\begin{figure}[htbp]
      \centering
      \includegraphics[width=\linewidth]{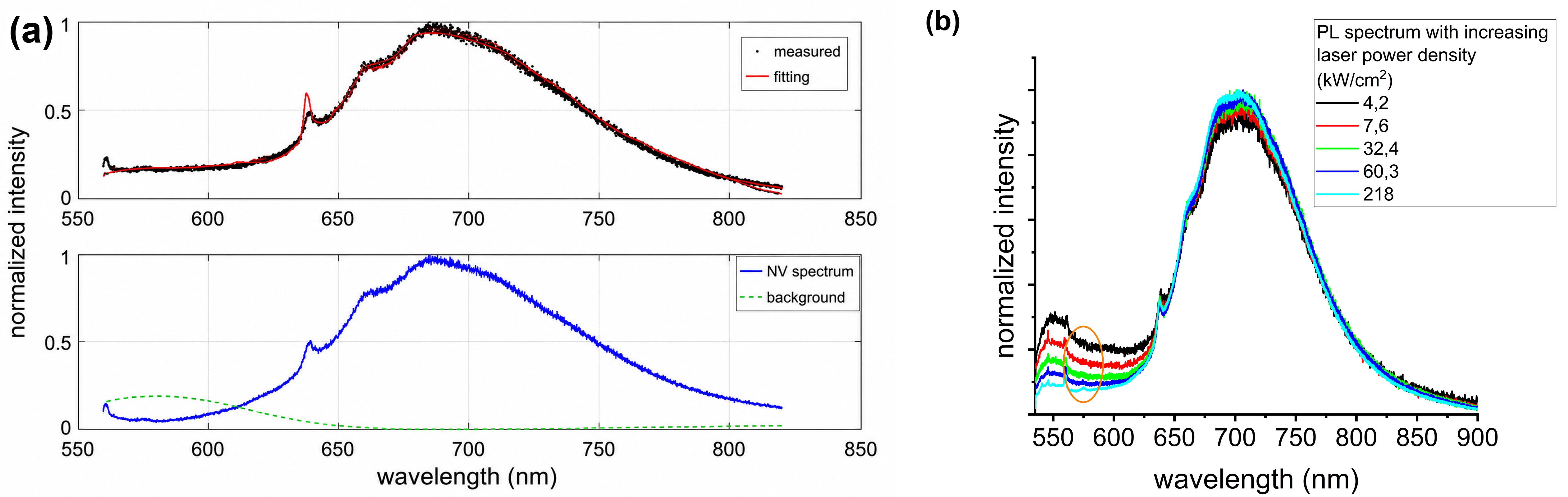}
      \caption{(a) Representative spectral decomposition according to Eq.~\eqref{eq:fit}. (b) Decrease of the luminescent background with increasing laser power.}
      \label{fig:plfit}
    \end{figure}

The PL spectrum of an NV-containing sample comprises the emission of both charge states, and their contributions were quantified following the decomposition procedure of Ref.~\cite{Rondin2010}, in which the spectrum, normalised to its total intensity, is expressed as a linear combination of the two charge-state components \cite{Thalassinos2025,Jegenyes2025,Czene2025}. Because the organic residue on sample~(2) also contributes to the PL spectrum, the model function was extended by a background term and reads

    \begin{equation}
      I_{\mathrm{fit}}(\lambda) = c_{-}\,S_{\ch{NV-}}(\lambda) + c_{0}\,S_{\ch{NV^0}}(\lambda) + G(\lambda) + B(\lambda),
      \label{eq:fit}
    \end{equation}

where $S_{\ch{NV-}}(\lambda)$ and $S_{\ch{NV^0}}(\lambda)$ are the reference emission spectra of the two charge states \cite{Alsid2019}, $c_{-}$ and $c_{0}$ are the corresponding weighting coefficients, $G(\lambda)$ is a Gaussian with its position fixed below \SI{600}{\nano\metre}, and $B(\lambda)$ accounts for the residual baseline and non-specific scattering. For the as-received material and ND\_XeF$_2$, no background component was required; setting $G(\lambda)=0$ reduces Eq.~\eqref{eq:fit} to the two-component decomposition of Refs.~\cite{Rondin2010,Thalassinos2025,Jegenyes2025,Czene2025}. The fits were performed in MATLAB. A representative fit is shown in Fig.~\ref{fig:plfit}(a); as shown in Fig.~\ref{fig:plfit}(b), the background component diminished with increasing laser power. From the fitted components, the \ch{NV-} fraction was calculated as

    \begin{equation}
      f_{\ch{NV-}} = \frac{I_{\ch{NV-}}}{I_{\ch{NV^0}} + I_{\ch{NV-}}},
      \label{eq:fraction}
    \end{equation}

where $I_{\ch{NV-}}$ and $I_{\ch{NV^0}}$ denote the integrated intensities of the respective weighted components.
 
\begin{figure}[htbp]
      \centering
      \includegraphics[width=\linewidth]{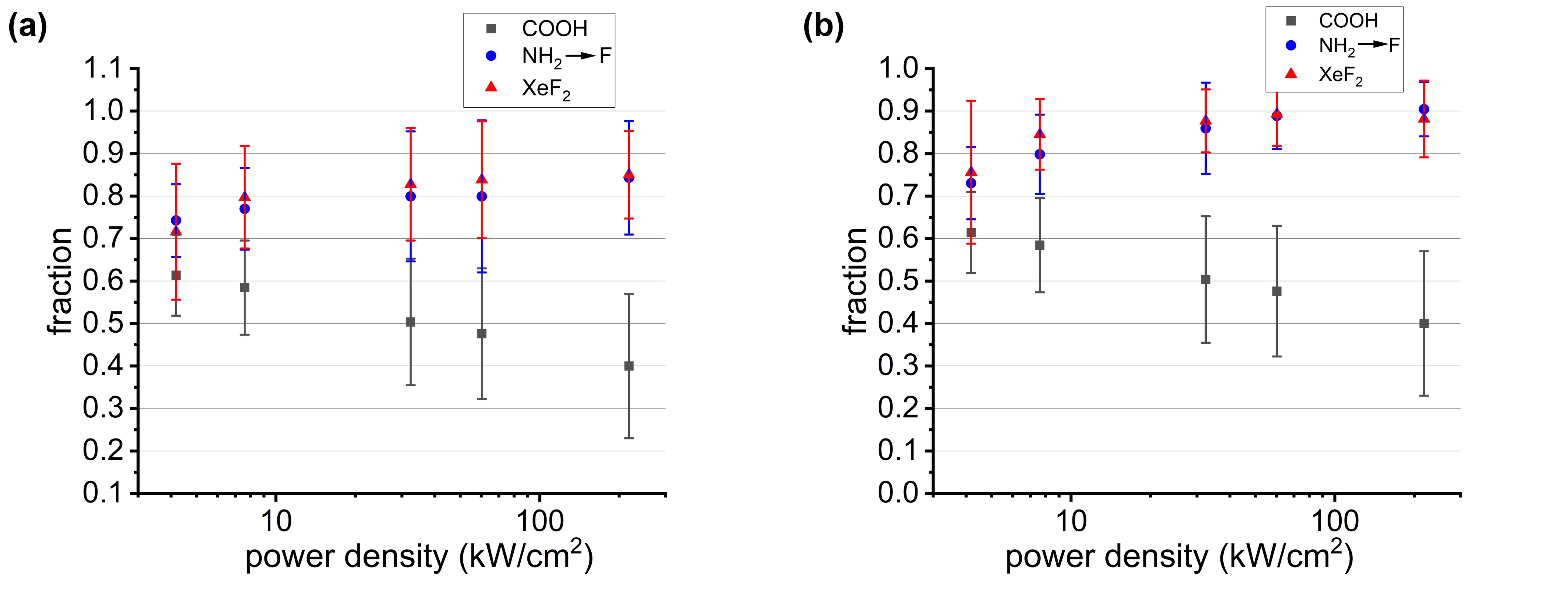}
      \caption{\ch{NV-} fraction as a function of laser power density for the fluorinated samples and the \ch{COOH}-terminated starting material. (a) Average of measurement points distributed across the films. (b) Measurements at the rim of the drop-cast films. The power density is averaged over the Airy disc of the focused laser spot (Section~2).}
      \label{fig:fraction}
    \end{figure}

The \ch{NV-} fractions (Eq.~\eqref{eq:fraction}) obtained for the two fluorinated samples and the as-received material are shown in Fig.~\ref{fig:fraction} and listed in Table~\ref{tab:fraction}. The data are presented in two panels: Fig.~\ref{fig:fraction}(a) shows measurement points distributed across the films, while Fig.~\ref{fig:fraction}(b) shows measurements at randomly chosen points along the rim of the drop-cast films, several hundred micrometres apart. The rim differs from the interior in film thickness and particle packing, and EDS shows that salt residues accumulate there (Section~3.1). The rim data are therefore shown separately and are not used to draw conclusions on the role of fluorine. At all applied powers, the \ch{NV-} fractions of the fluorinated samples exceeded those of the as-received material. The two data sets also exhibited opposite power dependences: in the as-received sample the fraction decreased monotonically with increasing power (from 0.61 to 0.40), whereas in the fluorinated samples it increased with power, from ${\approx}0.73$ to ${\approx}0.85$ (Fig.~\ref{fig:fraction}(a)). The two fluorination routes yield identical fractions within the measurement uncertainty at every power, although the fluorine coverage differs by a factor of about four. At the rim (Fig.~\ref{fig:fraction}(b) and Table~\ref{tab:fraction}), the fractions were higher still, saturating near 0.90, and individual points reached 0.9 even at the lowest applied power.

\paragraph{Residual adsorbates} Both routes leave residues on the surface, and their possible role in the stabilisation has to be considered. In ND\_XeF$_2$, pyridine is the only nitrogen source. The N\,1s spectrum contains no pyridinium component, so the HF/pyridine reagent itself is not retained, and the nitrogen content (Table~\ref{tab:xps}) limits the carbon in pyridine-derived species to ${\lesssim}\SI{5}{at.\percent}$ on average. This nitrogen persisted after the additional \ch{NaOH}/\ch{HCl} wash and is therefore strongly bound; it is also present in the material used for the optical measurements. ND\_XeF$_2$ further contains Cl, most likely from the \ch{CDCl3} solvent, whereas ND\_NH$_2$-F contains Br from the Hofmann step and the sp$^2$-rich carbonaceous by-product. The two routes thus leave different residues but give the same \ch{NV-} fractions, so the route-specific residues are unlikely to cause the stabilisation. Metal-fluoride deposits are found by EDS only at the rim of the drop-cast films and are minor in the areas analysed by XPS (Section~3.1), so they are unlikely to account for the stabilisation measured across the films.
 
    \begin{table}[htbp]
      \centering
      \caption{Applied laser powers and power densities (averaged over the Airy disc, Section~2) and the \ch{NV-} fractions measured at the rim of the films (Fig.~\ref{fig:fraction}(b)).}
      \label{tab:fraction}
      \begin{tabular}{ccccc}
        \toprule
        power (\si{\micro\watt}) & power density (\si{\kilo\watt\per\centi\metre\squared}) & \ch{COOH} & \ch{NH2 -> F} & \ch{XeF2} \\
        \midrule
        55.5 & 4.18 & 0.61 ($\pm$0.10) & 0.73 ($\pm$0.08) & 0.76 ($\pm$0.17) \\
        101  & 7.61 & 0.58 ($\pm$0.11) & 0.80 ($\pm$0.09) & 0.85 ($\pm$0.08) \\
        430  & 32.4 & 0.50 ($\pm$0.15) & 0.86 ($\pm$0.11) & 0.88 ($\pm$0.07) \\
        800  & 60.3 & 0.48 ($\pm$0.15) & 0.89 ($\pm$0.08) & 0.89 ($\pm$0.08) \\
        2900 & 218  & 0.40 ($\pm$0.17) & 0.90 ($\pm$0.06) & 0.88 ($\pm$0.09) \\
        \bottomrule
      \end{tabular}
    \end{table}

The photostability of the \ch{NV-} state was further examined through optically driven charge cycling. The two charge states coexist in a dynamic, illumination-dependent equilibrium \cite{Shinei2021,Aslam2013,Thalassinos2025}. Interconversion proceeds by two-photon processes in both directions: ionisation of \ch{NV-} occurs when a first photon excites the centre and a second photon promotes the excited electron to the conduction band, whereas recombination to \ch{NV-} requires excitation of \ch{NV^0} followed by electron capture from the valence band \cite{Aslam2013}. Under red illumination (\SI{633}{\nano\metre}), \ch{NV-} is still photoionised, but the reverse process is inefficient because \ch{NV^0} absorbs only weakly at this wavelength; prolonged red illumination therefore shifts the population towards \ch{NV^0}. Subsequent green illumination drives both processes and re-establishes the \ch{NV-}-rich steady state \cite{Aslam2013,Beha2012}. The extent to which the \ch{NV-} population is depleted during red illumination, and recovered afterwards, thus provides a measure of the charge-state photostability.
 
\begin{figure}[htbp]
      \centering
      \includegraphics[width=\linewidth]{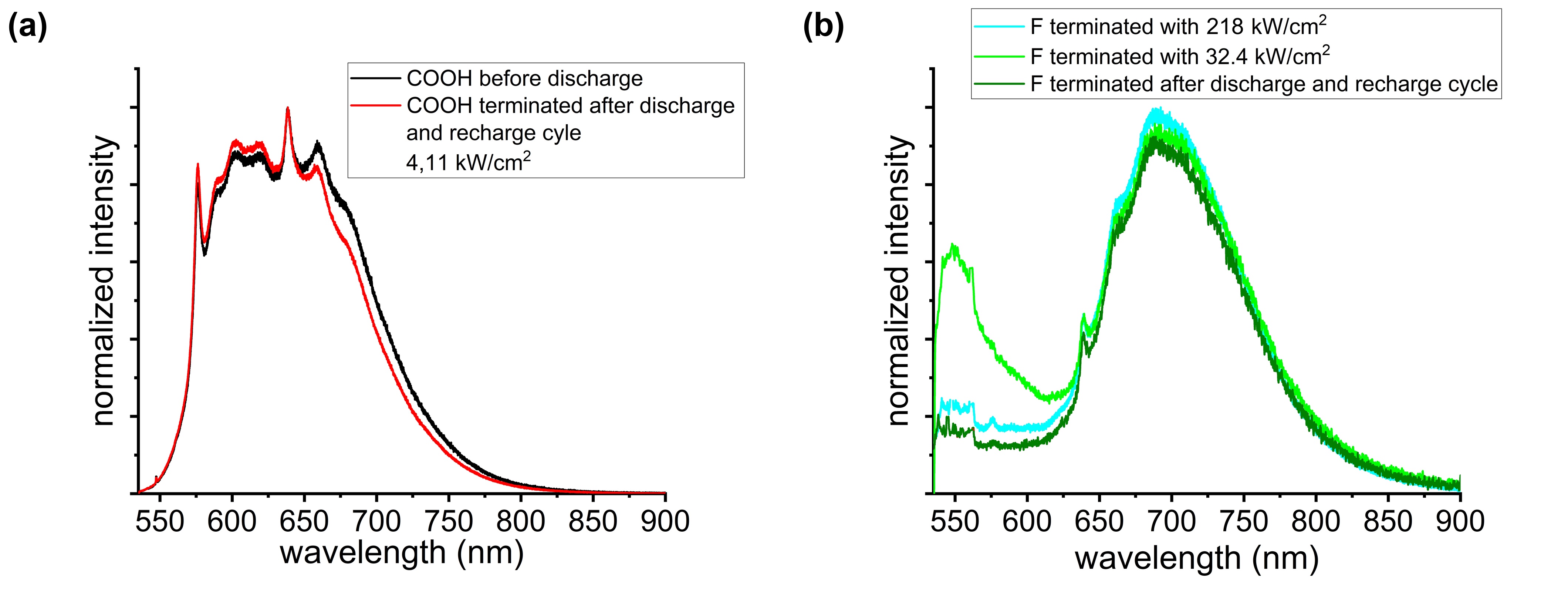}
      \caption{Discharge--recharge cycle of (a) the \ch{COOH}-terminated and (b) the fluorine-terminated sample.}
      \label{fig:discharge}
    \end{figure}

The PL spectra recorded before and after \SI{10}{\minute}  of \SI{633}{\nano\metre} illumination (\SI{77}{\micro\watt}, \SI{4.11}{\kilo\watt\per\centi\metre\squared} over the Airy disc of \SI{1.54}{\micro\metre}) are shown in Fig.~\ref{fig:discharge}. During this period of time, we followed the intensity change at the \ch{NV-} ZPL. In both cases, it decreased somewhat over time, but it never became completely 0. In the as-received sample, the \ch{NV-} fraction decreased from 0.15 to 0.07 upon the discharge--recharge cycle. Since oxygen-containing groups also possess a certain degree of stability effect, the \ch{NV-} centres may be more stable at those points or less stable, where the destabilising groups (e.g. COOH and H) are numerous. With this measurement, we wanted to demonstrate the high stability caused by the F termination. We compared this with a point in the as-received sample. In the fluorinated sample, by contrast, the \ch{NV-} fraction remained unchanged within the measurement uncertainty.
 
\paragraph{Spin properties: overview} The ODMR line positions, described by the zero-field splitting parameters $D$ and $E$, report on strain and static electric fields \emph{at} the NV centres.{} The spin--lattice relaxation time \ensuremath{T_1} reports on magnetic noise \emph{around} them, which in nanodiamonds originates mainly at the surface.{} The main result is that $D$ and $E$ are unchanged by fluorination, so the NV centres and the diamond lattice are not modified{}, whereas \ensuremath{T_1} is shortened by about \SI{40}{\percent}, which shows that the \ch{NV-} population now includes centres close to the surface{}. Together with the PL data, this identifies the effect of fluorination as an indirect, surface-electrostatic one.

\begin{figure}[htbp]
      \centering
      \includegraphics[width=\linewidth]{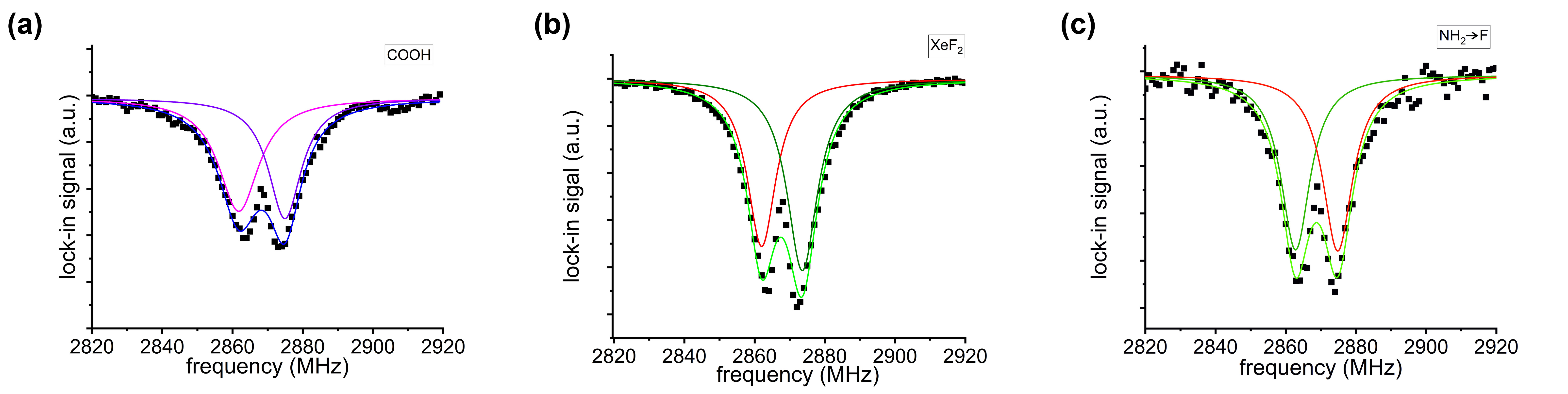}
      \caption{ODMR spectra of (a) the \ch{COOH}-terminated, (b) the \ch{XeF2}-treated, and (c) the BS-treated (\ch{NH2 -> F}) NDs. The applied microwave power was \SI{-36}{dBm} and the laser power \SI{220.5}{\milli\watt}. The coloured lines are the fitted curves.}
      \label{fig:odmr}
    \end{figure}

\begin{figure}[htbp]
      \centering
      \includegraphics[width=0.7\linewidth]{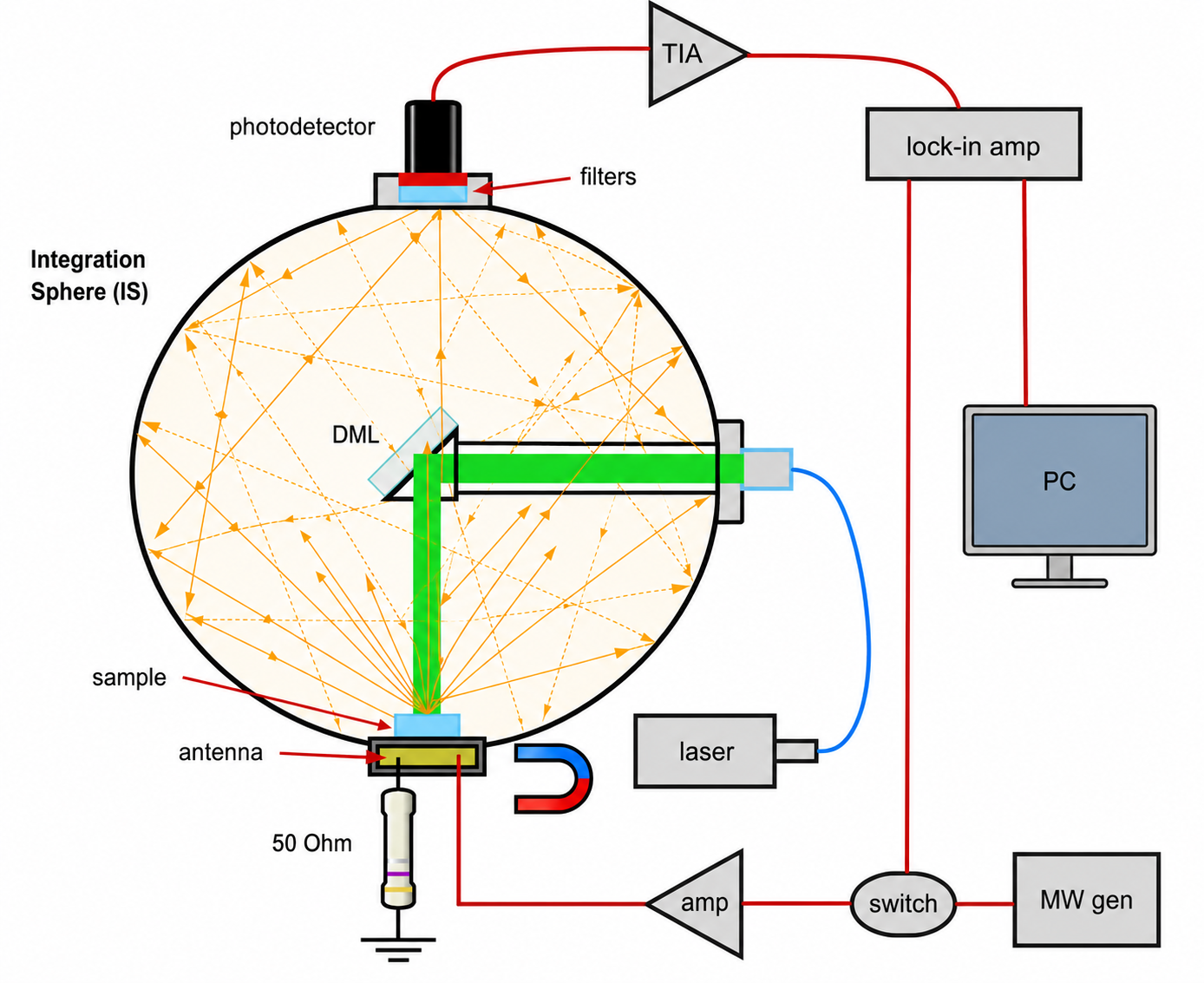}
      \caption{Schematic of the ODMR measurement setup.}
      \label{fig:setup}
    \end{figure}

\paragraph{ODMR} The spin properties of the centres were characterised by ODMR. From CW-ODMR measurement the zero-field splitting parameters $D$ and $E$ were extracted. Besides the axial parameter $D$, which originates from the electron spin--spin interaction, the transverse parameter $E$ is often nonzero in diamond NV centers and sensitive to transverse strain and to local electric fields, including those of nearby charges \cite{Doherty2013,Dolde2011,Mittiga2018}. Together, the two parameters therefore probe whether fluorination alters the crystalline and electrostatic environment of the \ch{NV-} centres. The ODMR spectra are shown in Fig.~\ref{fig:odmr}; a schematic of the measurement system is given in Fig.~\ref{fig:setup}. At zero applied magnetic field, the ensemble spectra exhibit two resonances at $\nu_{\pm} = D \pm E$, and the spectra were fitted accordingly (coloured lines in Fig.~\ref{fig:odmr}). The resulting $D$ and $E$ values are listed in Table~\ref{tab:de}.
 
    \begin{table}[htbp]
      \centering
      \caption{$D$ and $E$ parameters (\si{\mega\hertz}) determined for the \ch{COOH}-terminated and fluorinated NDs.}
      \label{tab:de}
      \begin{tabular}{lccc}
        \toprule
        & \ch{COOH} & \ch{XeF2} & \ch{NH2 -> F} \\
        \midrule
        $D$ & 2868.8 ($\pm$0.1) & 2869.1 ($\pm$0.2) & 2868.8 ($\pm$0.2) \\
        $E$ & 6.2 ($\pm$0.1)    & 6.2 ($\pm$0.2)    & 6.0 ($\pm$0.2)    \\
        \bottomrule
      \end{tabular}
    \end{table}
 
Both parameters were unchanged within the measurement uncertainty across the three samples. The invariance of $D$ indicates that neither treatment introduced measurable axial strain, and the invariance of $E$ that the ensemble-averaged transverse strain and quasi-static electric-field environment of the centres is likewise unaffected. Fluorination therefore does not act on the NV centres directly; it changes which of them are in the negative charge state. 

\subsection{Spin--lattice relaxation ($T_1$)}
\begin{figure}[htbp]
      \centering
      \includegraphics[width=\linewidth]{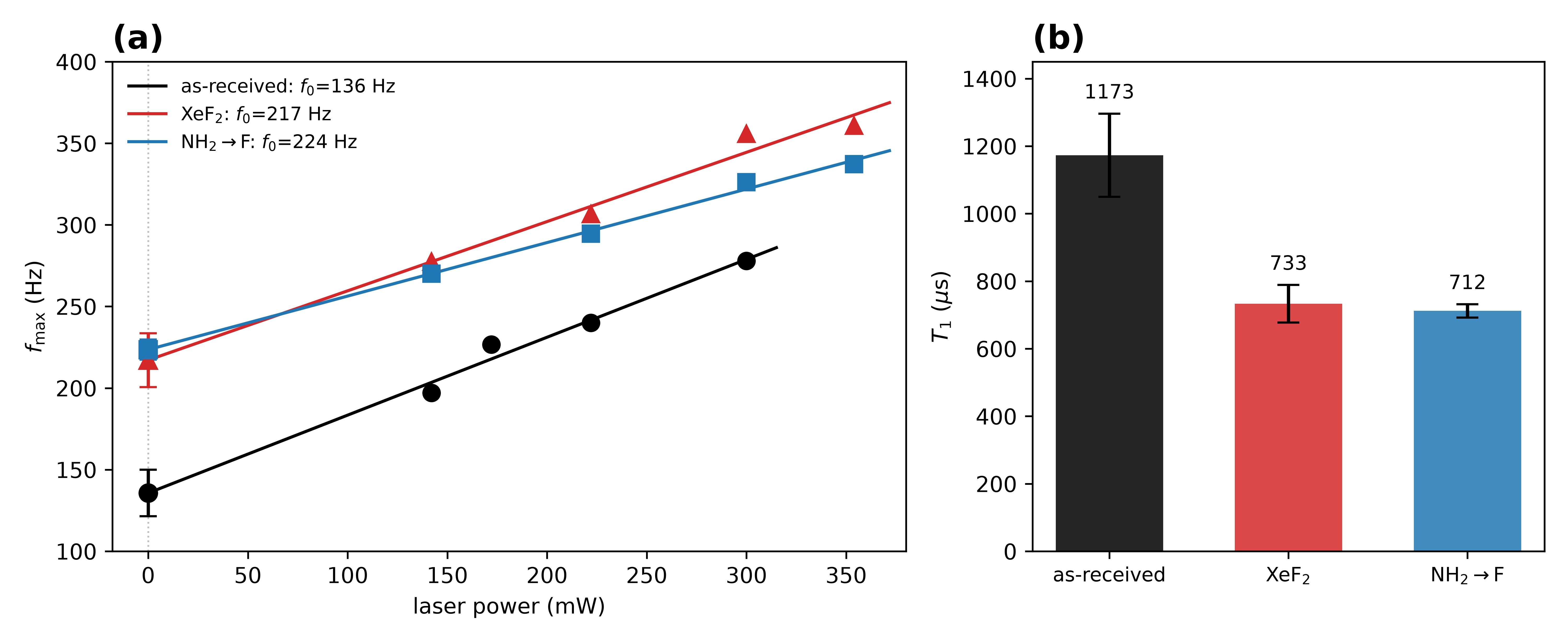}
      \caption{(a) Position of the maximum of the lock-in response, $f_{\mathrm{max}}$, as a function of laser power for the three samples, with linear fits extrapolated to zero laser power; the filled symbols on the ordinate are the intercepts $f_{0}$ with their standard errors. The as-received point at \SI{222}{\milli\watt} is the mean of two repeated measurements. (b) The resulting spin--lattice relaxation times $T_1 = 1/(2\pi f_{0})$. All data were recorded at \SI{25}{\dBm} microwave power at the sample.}
      \label{fig:t1fit}
    \end{figure}

The ensemble spin--lattice relaxation time \ensuremath{T_1} was measured on the same particles by CW frequency-domain relaxometry~\cite{Verkhovlyuk2026}, using the ODMR setup described in the Methods; the results are summarised in Fig.~\ref{fig:t1fit} and Table~\ref{tab:t1}. Each value is an ensemble average over the very large number of nanoparticles contained in the illuminated spot, so that the reported \ensuremath{T_1} characterises the particle population as a whole rather than individually selected crystals. The as-received, \ch{COOH}-terminated particles show $T_1 = 1173\pm123$~\si{\micro\second}, of the same order as the value ($\sim$\SI{1045}{\micro\second}) reported for HPHT FNDs of comparable size~\cite{Alkahtani2025}. Contrary to what the charge-state data alone might suggest, both fluorination routes \emph{shortened} \ensuremath{T_1}, to $733\pm56$~\si{\micro\second} for ND\_XeF$_2$ and $712\pm20$~\si{\micro\second} for ND\_NH$_2$-F, a reduction of about \SI{40}{\percent} in both cases. The two routes agree with each other within the measurement uncertainty, mirroring their behaviour in the \ch{NV-} fraction despite the four-fold difference in fluorine coverage (Table~\ref{tab:xps}).
The opposite signs of the two effects are not contradictory. Fluorination stabilises \ch{NV-} throughout the particle, including centres in the near-surface region that previously remained neutral; these are also the centres most strongly coupled to the residual surface spin bath, so that their recruitment raises the measured \ch{NV-} fraction while lowering the ensemble-averaged \ensuremath{T_1}. The adsorbed organic layer seen by FTIR and Raman, and the increased sp$^2$ content seen by XPS, provide further relaxation channels in the same near-surface region. The values reported here accordingly remain below those reached for nanodiamonds of comparable size by treatments that remove or bury the surface spin bath rather than repopulate \ch{NV-}{}---chemical-vapour-deposition growth combined with electron irradiation and high-temperature annealing (\SI{4.7}{\milli\second} in \SI{100}{\nano\metre} CVD NDs) \cite{Mameli2026}, molten-\ch{KNO3} oxidation ($\sim$\SI{2}{\milli\second}) \cite{Alkahtani2025}, or silica over-coating ($\sim$\SI{1}{\milli\second}) \cite{Barzegar2025}.

Taking the increase of the film-averaged \ch{NV-} fraction at the lowest laser power ($0.61 \rightarrow 0.72$--0.74, Fig.~\ref{fig:fraction}(a)) and assuming a uniform NV distribution, the converted population corresponds to an equivalent surface shell $d_{\mathrm{eff}} = \Delta f\,(V/A) \approx 2.5$--\SI{3}{\nano\metre}, where $V/A = d/6$ is the volume-to-surface ratio of a spherical particle of diameter $d = \SI{140}{\nano\metre}$. The spherical approximation used here gives an upper bound, since any faceted particle has a larger surface-to-volume ratio and hence a thinner implied shell. This is of the same order as the few-nanometre depth over which upward band bending is expected to deplete \ch{NV-} near the surface~\cite{Hauf2011,broadway2018}, and is therefore consistent with the near-surface origin proposed above. We stress that this is an order-of-magnitude consistency check rather than a determination of the shell thickness: it assumes a uniform NV distribution and treats the photoluminescence-weighted \ch{NV-} fraction as a number fraction.

The relaxation times retained after fluorination nevertheless remain well suited to relaxometry. Values of ${\sim}\SI{0.7}{\milli\second}$ are several times longer than the ${\sim}\SI{0.17}{\milli\second}$ reported for commercial HPHT fluorescent nanodiamonds \cite{Mameli2026}, and lie comfortably within the range exploited for the detection of free radicals and paramagnetic ions \cite{Fan2025,Nie2021,Steinert2013}. More importantly, because the relaxation rate induced by an external target falls off as $r^{-6}$ \cite{Tetienne2013,Schirhagl2014}, the relaxometric signal is dominated by the NV centres lying closest to the surface. Fluorination converts precisely these centres from the sensing-inactive \ch{NV^0} into \ch{NV-}, and thereby increases the number of usable sensors in the outermost shell of the particle, where the coupling to an analyte is strongest. On this view the moderate reduction of the ensemble-averaged \ensuremath{T_1} is the expected signature of having recruited this near-surface population, and is traded against a larger effective sensing volume rather than against sensing capability as such. 
 
    \begin{table}[htbp]
      \centering
      \caption{Ensemble NV spin--lattice relaxation time \ensuremath{T_1} for the as-received and F-terminated samples, obtained by extrapolating the measured relaxation rates to zero laser power. All three samples were measured under identical microwave conditions (\SI{25}{\dBm} at the sample). The quoted uncertainties are the standard errors of the zero-power extrapolation.}
      \label{tab:t1}
      \begin{tabular}{lc}
        \toprule
        Sample & \ensuremath{T_1} (\si{\micro\second}) \\
        \midrule
        \ch{COOH} (as-received) & $1173 \pm 123$ \\
        \ch{XeF2}               & $733 \pm 56$ \\
        \ch{NH2 -> F} (BS)      & $712 \pm 20$ \\
        \bottomrule
      \end{tabular}
    \end{table}

Because the converted shell has a fixed thickness, its share of the particle volume, and with it the relative effect of fluorination, grows as the particles become smaller (Table~\ref{tab:size}). Smaller particles are therefore preferable for sensing that requires close proximity to the analyte. The \SI{140}{\nano\metre} particles were chosen here because each hosts several hundred NV centres. This makes the \ch{NV-} fraction obtained from the PL decomposition statistically well defined and gives the ensemble ODMR and relaxometry measurements adequate signal-to-noise. Below ${\sim}\SI{50}{\nano\metre}$, a particle hosts only a few to a few tens of NV centres at the same concentration, and the \ch{NV-} fraction fluctuates strongly from particle to particle, which would make the comparison between surface terminations much less precise.

    \begin{table}[htbp]
      \centering
      \caption{Number of NV centres per particle (\SI{3}{\ppm} NV) and volume fraction of a \SI{3}{\nano\metre} near-surface shell for spherical particles of different diameter.}
      \label{tab:size}
      \begin{tabular}{ccc}
        \toprule
        Diameter (\si{\nano\metre}) & NV centres per particle & Shell volume fraction \\
        \midrule
        140 & ${\approx}760$ & 0.12 \\
        100 & ${\approx}280$ & 0.17 \\
        50  & ${\approx}35$  & 0.32 \\
        20  & ${\approx}2$   & 0.66 \\
        \bottomrule
      \end{tabular}
    \end{table}

\section{Conclusions}
This work set out to improve the quantum-optical properties of NV centres in nanodiamond by mild surface fluorination while preserving the integrity of the diamond lattice. Two solution-phase routes were developed and compared: a Balz--Schiemann-type reaction, adapted here to convert amino-terminated nanoparticles into \ch{C-F}-terminated ones, in which HF serves as the fluoride source---a highly toxic reagent, yet one far easier to handle than elemental \ch{F2} or \ch{ClF3}---and the second involved \ch{XeF2} in the solution phase, where hydrofluoric acid (HF) acts both as a catalyst and as a solvent for \ch{XeF2}, as described in the literature. The formation of \ch{C-F} bonds was established by XPS and is supported by FTIR, with fluorine contents of \SI{2.6}{\percent} (\ch{XeF2}) and \SI{0.7}{\percent} (BS), corresponding to sub-monolayer coverage. Raman spectroscopy showed that the diamond lattice is preserved in both routes. Both synthesis routes provided equal \ch{NV-} stabilisation, with the \ch{NV-} fraction reaching ${\approx}0.85$ on average and ${\approx}0.9$ at the rim of the films, and photostability upon prolonged \SI{633}{\nano\metre} illumination{}; the spin--lattice relaxation time, by contrast, was shortened by about \SI{40}{\percent} by both treatments, while remaining near \SI{0.7}{\milli\second} and thus well within the range useful for relaxometric sensing{}.
 
Notably, the two routes yield the same charge-state behaviour within the measurement uncertainty despite a roughly four-fold difference in fluorine coverage, indicating again that mixed surface termination already suffices for stabilisation. We attribute the stabilisation to the combined effect of removing acceptor-like carboxyl groups, the positive electron affinity of the \ch{C-F} termination, and the thinner adsorbed water layer on the more hydrophobic surface. We assume that the dominant source of the charge-state stabilisation is the \ch{C-F} termination of the nanodiamond surface. {}The two routes also yield the same spin--lattice relaxation time within uncertainty, so the shortening of \ensuremath{T_1} appears to be a generic consequence of the fluorinated surface rather than of the carbonaceous residue specific to the BS product. Because the relaxometric response to an external spin falls off as $r^{-6}$, the near-surface centres that fluorination converts into \ch{NV-} are precisely those that dominate sensing, so the shorter ensemble \ensuremath{T_1} is better read as the cost of activating a larger and more strongly coupled sensor population than as a loss of performance; direct relaxometric measurements on an external target would be required to confirm that this trade is favourable in practice. {}Although not yet fully optimised, the BS method already matches the \ch{XeF2} treatment in charge-state performance using bench-scale diazonium chemistry, and offers clear room for improvement through refined purification and reaction control. A remaining consideration is the increased hydrophobicity of fluorine-terminated surfaces, which lowers colloidal stability in aqueous media and is therefore relevant for biological applications; it can, however, be mitigated by secondary functionalisation or mixed terminations, and it is of little consequence for dry and solid-state sensing configurations. Because the converted near-surface shell has a fixed thickness of a few nanometres, the relative effect of fluorination is expected to grow for smaller particles. Taken together, these results establish mild, solution-phase fluorination as a simple and scalable route to charge-stable NV ensembles in nanodiamond, and motivate its extension to shallow NV centres in bulk diamond, where surface noise is most limiting.

\section*{CRediT authorship contribution statement}
\textbf{Szabolcs Czene:} Conceptualization, Methodology, Writing -- original draft.
\textbf{Olga Krafcsik:} Methodology, Writing -- review \& editing.
\textbf{Nikoletta Jegenyés:} Methodology, Writing -- original draft.
\textbf{Dávid Beke:} Writing -- review \& editing.
\textbf{László Péter:} Methodology, Writing -- review \& editing.
\textbf{Vladimir Verkhovlyuk:} Methodology, Writing -- original draft.
\textbf{Ádám Gali:} Conceptualization, Funding acquisition, Supervision, Writing -- review \& editing.

\section*{Declaration of competing interest}
The authors declare that they have no known competing financial interests or personal relationships that could have appeared to influence the work reported in this paper.

\section*{Data availability}
The data supporting this study (raw XPS, FTIR, Raman, EDS, photoluminescence, ODMR and relaxometry data) have been deposited in the HUN-REN data repository and will be made publicly available under a DOI, which will be added at the proof stage. Until then, the data are available from the corresponding author on request.

\section*{Acknowledgements}
{}The authors thank Anton Pershin for valuable advice on the frequency-domain relaxometry measurements. {}The authors thank Katalin Kamarás, Erzsébet Takács and Elemér Vass for valuable discussions on the interpretation of the FTIR spectra. This work was supported by the EUREKA project quNV2.0 (2021-17465/NP/BILAT\_HU\_DE\_1, 2023--2026) via the National Research, Development and Innovation Office (NKFIH) under Grant No.\ 2020-1.2.3-EUREKA-2022-00022. The research reported in this paper and carried out at the Wigner Research Centre for Physics is supported by the infrastructure of the Hungarian Academy of Sciences.

\section*{Declaration of generative AI and AI-assisted technologies in the manuscript preparation process}
During the preparation of this work the authors used Claude (Anthropic), accessed through Claude Code, in order to generate the graphical abstract (plotted from the measured data with AI-generated Python code), to assist in writing the data-analysis and plotting scripts (Section~2), and to draft and edit parts of the text under the full control of the corresponding author. After using this tool, the authors reviewed and edited the content as needed and take full responsibility for the content of the published article.

\bibliographystyle{elsarticle-num}
\bibliography{references}

\appendix
\section{Frequency-domain relaxometry spectra}
\setcounter{table}{0}\setcounter{figure}{0}
Figure~\ref{fig:si_t1spectra} shows the CW-ODMR lock-in response as a function of the microwave amplitude-modulation frequency for the three samples at each of the laser powers used for the \ensuremath{T_1} determination. In every sample the maximum shifts to higher modulation frequency as the laser power is increased, reflecting the additional, optically induced contribution to the measured relaxation rate; this power dependence is what the extrapolation to zero laser power removes. Between \num{14} and \num{35} scans were averaged per trace, and the sweeps start above \SI{35}{\hertz} so as to exclude the narrow interference spikes that dominate the response at lower modulation frequencies. The position of the maximum was obtained by fitting a parabola to the response in $\log_{10} f$ over $\pm 0.35$ decades about the maximum; the resulting $f_{\mathrm{max}}$ is stable to within a few Hz against the choice of this window.

\begin{figure}[htbp]
  \centering
  \includegraphics[width=\linewidth]{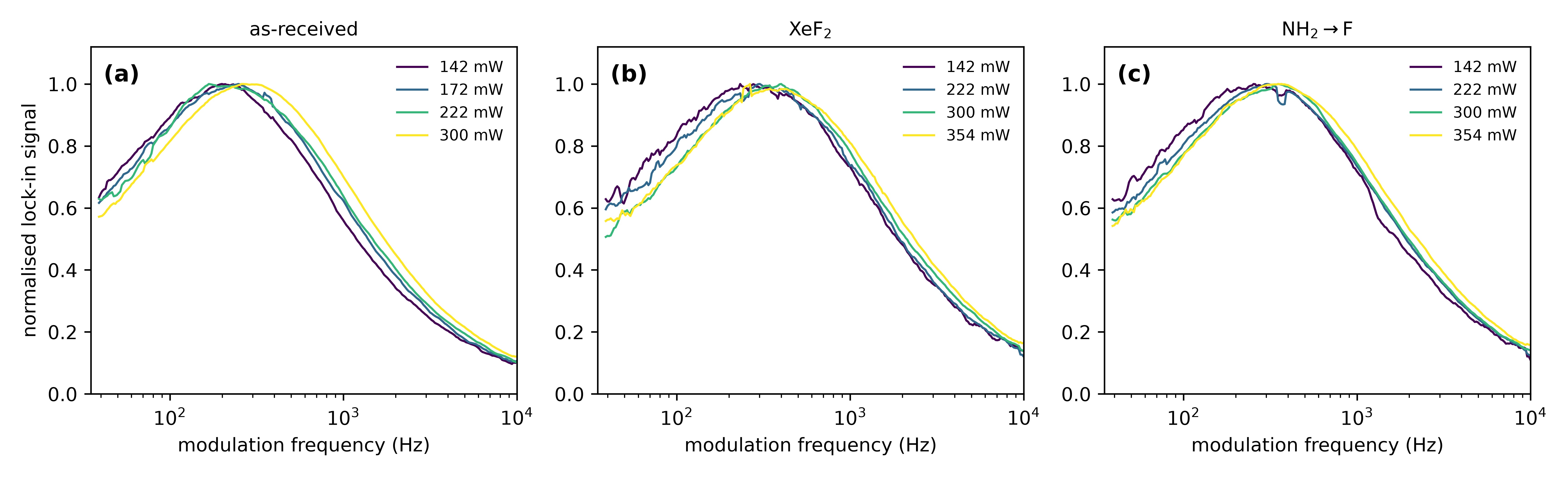}
  \caption{Lock-in signal versus microwave amplitude-modulation frequency for (a) the as-received \ch{COOH}-terminated, (b) the \ch{XeF2}-treated, and (c) the BS-treated (\ch{NH2 -> F}) nanodiamonds, at the laser powers used for the \ensuremath{T_1} determination and a microwave power of \SI{25}{\dBm} at the sample. Each trace is normalised to its own maximum.}
  \label{fig:si_t1spectra}
\end{figure}

\section{FTIR substrate controls}
\label{app:ftir}
\setcounter{table}{0}\setcounter{figure}{0}

\begin{table}[htbp]
  \centering
  \caption{Assignment of the FTIR bands discussed in Section~3.1.}
  \label{tab:ftir}
  \small
  \begin{tabular}{p{0.15\linewidth}p{0.45\linewidth}p{0.3\linewidth}}
    \toprule
    Band (\si{\per\centi\metre}) & Assignment & Occurrence \\
    \midrule
    1263 & \ch{C-F} stretching on sp$^3$ carbon (tentative)~\cite{Jeon2011,Ueta2012c,Ando1995,Osipov2020,Zagrebina2015} & fluorinated samples only; follows the XPS fluorine content \\
    ${\sim}1340$ & \ch{C-F} of extensively fluorinated nanodiamond~\cite{Liu2004,Osipov2020} & not observed \\
    ${\sim}1365$ & not assigned; nitrogen-containing groups or carbonaceous by-product & ND\_NH$_2$-F only \\
    1107 & interstitial oxygen of the Si substrate, possibly with a \ch{C-F} contribution~\cite{Liu2004,Zagrebina2015} & all samples; not quantifiable \\
    1658 & C=O stretching of carbonyl groups, adsorbed water~\cite{Petit2018,Coates2006} & all samples \\
    ${\sim}800$ & not related to fluorination & all samples \\
    $>1800$ & C(O)F groups would absorb here & not observed \\
    611 & Si lattice band & substrate \\
    \bottomrule
  \end{tabular}
\end{table}

\begin{figure}[htbp]
  \centering
  \includegraphics[width=\linewidth]{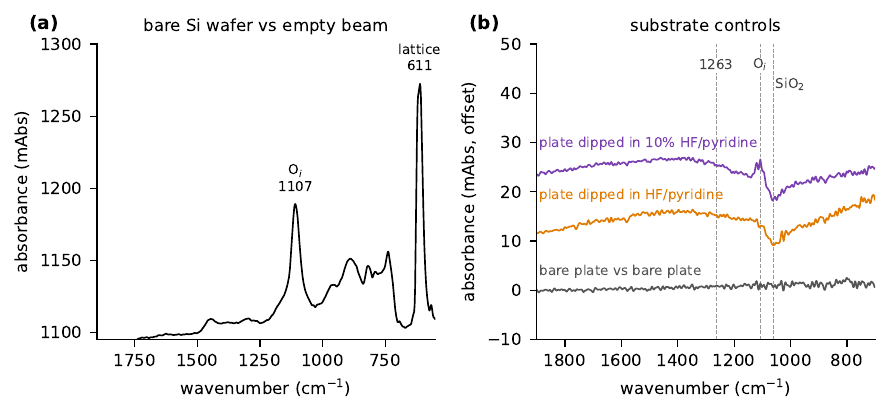}
  \caption{FTIR substrate controls (IFS~66v/S, under vacuum). (a) Bare Si wafer piece measured against the empty beam, showing the lattice band at \SI{611}{\per\centi\metre} and the interstitial-oxygen band at \SI{1107}{\per\centi\metre}. (b) Bare wafer pieces measured against a bare wafer piece: untreated, and after dipping into concentrated and into 10\% HF/pyridine (offset for clarity). The dashed lines mark \SI{1263}{\per\centi\metre}, the O$_i$ band and the broad \ch{SiO2} band of the native oxide.}
  \label{fig:si_ftir}
\end{figure}

The nanodiamond films were drop-cast onto pieces of one Czochralski Si wafer. Figure~\ref{fig:si_ftir}a shows the absorbance of a bare wafer piece against the empty beam. The two bands relevant here are the lattice band at \SI{611}{\per\centi\metre} and the interstitial-oxygen (O$_i$) band at \SI{1107}{\per\centi\metre}. Figure~\ref{fig:si_ftir}b shows control measurements against a bare wafer piece as background:
\begin{itemize}
  \item a second bare piece, which gives a flat spectrum;
  \item pieces dipped into concentrated and into diluted (10\%) HF/pyridine, the reagent used in the syntheses.
\end{itemize}
The empty beam measured against the empty beam was flat to within \SI{0.1}{mAbs}, which excludes contributions of the light source. The HF/pyridine treatment removes part of the native oxide, which appears as a broad negative band near \SI{1060}{\per\centi\metre}, but produces no band near \SI{1263}{\per\centi\metre}. The narrow feature at \SI{1107}{\per\centi\metre} on the dipped pieces has the position and width of the O$_i$ band. It reflects the slightly different oxygen content of different wafer pieces. This is why the sample contribution near \SI{1107}{\per\centi\metre} cannot be quantified.

\end{document}